\documentclass[aasmacros]{aa}

\usepackage{amsmath}
\usepackage{wasysym}
\usepackage{marvosym}
\usepackage{txfonts}
\usepackage{mathtools}
\usepackage{placeins}
\usepackage{siunitx}
\usepackage{enumitem}
\DeclareSIUnit[]\astronomicalunit{\text{au}}

\usepackage{natbib}
\bibpunct{(}{)}{;}{a}{}{,} 

\usepackage{epstopdf}
\usepackage{graphicx}
\pdfoutput=1
\usepackage[pdftex,dvipsnames]{color,xcolor}
\usepackage{subcaption}

\usepackage{scalerel}
\usepackage{tikz}
\usetikzlibrary{svg.path}

\definecolor{orcidlogocol}{HTML}{A6CE39}
\tikzset{
  orcidlogo/.pic={
    \fill[orcidlogocol] svg{M256,128c0,70.7-57.3,128-128,128C57.3,256,0,198.7,0,128C0,57.3,57.3,0,128,0C198.7,0,256,57.3,256,128z};
    \fill[white] svg{M86.3,186.2H70.9V79.1h15.4v48.4V186.2z}
                 svg{M108.9,79.1h41.6c39.6,0,57,28.3,57,53.6c0,27.5-21.5,53.6-56.8,53.6h-41.8V79.1z M124.3,172.4h24.5c34.9,0,42.9-26.5,42.9-39.7c0-21.5-13.7-39.7-43.7-39.7h-23.7V172.4z}
                 svg{M88.7,56.8c0,5.5-4.5,10.1-10.1,10.1c-5.6,0-10.1-4.6-10.1-10.1c0-5.6,4.5-10.1,10.1-10.1C84.2,46.7,88.7,51.3,88.7,56.8z};
  }
}

\newcommand\orcidicono[1]{\href{https://orcid.org/0000-0001-6693-7910/#1}{\mbox{\scalerel*{
\begin{tikzpicture}[yscale=-1,transform shape]
\pic{orcidlogo};
\end{tikzpicture}
}{|}}}}

\newcommand\orcidiconc[1]{\href{https://orcid.org/0000-0002-1013-2811/#1}{\mbox{\scalerel*{
\begin{tikzpicture}[yscale=-1,transform shape]
\pic{orcidlogo};
\end{tikzpicture}
}{|}}}}

\newcommand\orcidicona[1]{\href{https://orcid.org/0000-0002-8811-1914/#1}{\mbox{\scalerel*{
\begin{tikzpicture}[yscale=-1,transform shape]
\pic{orcidlogo};
\end{tikzpicture}
}{|}}}}

\newcommand\orcidiconr[1]{\href{https://orcid.org/0000-0001-5555-2652/#1}{\mbox{\scalerel*{
\begin{tikzpicture}[yscale=-1,transform shape]
\pic{orcidlogo};
\end{tikzpicture}
}{|}}}}

\usepackage[draft=False, colorlinks=true, citecolor=blue, urlcolor=blue, linkcolor=red, linktoc=all]{hyperref}

\usepackage{pifont}

\DeclareMathOperator\erf{erf}

\def\mearth{M_\oplus}

\def\msun{M_\odot}

\def\f1{f_{\rm I}}
\def\mstar{M_*}
\def\rstar{R_*}

\def\beq{\begin{equation}}
\def\eeq{\end{equation}}

\def\t2{\tau_{\rm II}}

\def\sigmas0{\Sigma_{\rm s,0}}
\def\mj{M_{\textrm{\tiny \jupiter }}}

\newcommand{\rj}{R_{\textrm{\tiny \jupiter}}}

\def\s0{S_0}

\def\apjl{ApJ}                
\def\apjs{ApJS}               
\def\ao{Appl.Optics}          
\def\aap{A\&A}                

\def\({\left(}
\def\){\right)}
\def\<{\left<}
\def\>{\right>}

\defcitealias{2021A&A...645A..43S}{S21}
\defcitealias{2022A&A...664A.138S}{S22}
\defcitealias{2023A&A...669A..31S}{S23}
\defcitealias{2008ApJ...685..560D}{DL08}
\defcitealias{2018MNRAS.475.5618B}{B18}
\defcitealias{Schib2025b}{Paper~II}

\begin{document}

\title{DIPSY: A new \textbf{D}isc \textbf{I}nstability \textbf{P}opulation \textbf{SY}nthesis}
\subtitle{I. Modeling, evolution of individual systems, and tests}

    \author{O. Schib\inst{\ref{WP}\orcidicono{}}
          \and
          C.~Mordasini\inst{\ref{WP},\ref{CSH}\orcidiconc{}}
          \and
          A.~Emsenhuber\inst{\ref{WP}\orcidicona{}}
          \and
          R.~Helled\inst{\ref{uzh}\orcidiconr{}}
          }

   \institute{
Space Research and Planetary Sciences, Physics Institute, University of Bern, Gesellschaftsstrasse 6, 3012 Bern, Switzerland\\
\email{oliver.schib@unibe.ch}
\label{WP}
\and
Center for Space and Habitability, University of Bern, Gesellschaftsstrasse 6, 3012 Bern, Switzerland\\
\label{CSH}
\and
Department of Astrophysics,
Universit\"at Z\"urich, Winterthurerstrasse~190, 8057~Z\"urich, Switzerland\\
\label{uzh}
             }

\date{Received July 4, 2025 / Accepted September 10, 2025}

\abstract
{Disc instability (DI) is a model aimed at explaining the formation of companions through the fragmentation of the circumstellar gas disc.
Furthermore, DI could explain the formation of part of the observed exoplanetary population.
Our understanding of DI is still incomplete given the complex nature of the process and  challenges related to its modelling.}
{We aim to provide a new comprehensive global model for the formation of companions via DI.
The model includes the formation of the star-and-disc system through infall from the molecular cloud core (MCC) as well as the evolution of companions that might emerge via fragmentation in it.
This approach allows us to study a large parameter space, perform population synthesis calculations, and make predictions that can be compared with observations. This makes it possible to put the models for the different sub-processes included in the global model to the observational test.}
{We have developed a global formation and evolution model for companions formed via DI: DIPSY. 
The model solves the 1D vertically integrated viscous evolution equation of the protostellar and protoplanetary disc with a variable $\alpha$ viscosity. It includes infall from the MCC, stellar irradiation, internal and external photoevaporation, and exchange of both mass and angular momentum between disc and companions. The latter leads for the companions to orbital migration and damping of the eccentricities and inclinations. 
As it evolves, the disc is continuously monitored for self-gravity and fragmentation.
When the conditions are satisfied, one (or several) clumps are inserted.
The evolution of the clumps is then followed in detail.
Clump contraction, including the second collapse is considered by using interior evolution tracks. Thermal irradiation by the disc, mass growth by gas accretion, and mass loss via Roche lobe overflow are also included.
The interaction of clumps with each other is included with a full N-body integrator which can lead to collisions, scattering, or ejections.}
{We showcased the model by performing a number of simulations for various initial conditions, from simple non-fragmenting systems to complex systems with many fragments.} 
{We confirm that the DIPSY model is a comprehensive and versatile global model of companion formation via DI.  It  enables  studies of the formation of companions with planetary to low stellar masses  around primaries with final masses that range from the brown dwarf to the B-star regime. 
We conclude that it is necessary to consider the many interconnected processes such as gas accretion, orbital migration, and N-body interactions, as they strongly influence the inferred population of forming objects. It is also clear that model assumptions play a key role in the determination of the systems undergoing formation.}


\maketitle

\section{Introduction}\label{sect:Introduction}
Planet formation occurs in discs around young stars and two main pathways for this process have been proposed: core accretion (CA, \citealt{safronov1972,Pollack1996,2005Icar..179..415H,Alibert2005}) and disc instability (DI, \citealt{1951PNAS...37....1K,1978PThPh..60..699M,Cameron1978,1997Sci...276.1836B,Boss2003}).
In the case of DI (also known as gravitational instability, GI), a circumstellar disc fragments under its own gravity to form one or several gravitationally bound objects.
These so called `clumps' have masses of the order of \num{1} Jovian mass, sizes of $\sim$~\SI{1}{au} (astronomical unit), and they are close to hydrostatic equilibrium.
The subsequent evolution of the clumps determines whether they survive and  become compact companions. 
Companions formed this way can lie in the planetary, brown dwarf, or stellar mass regime.
This includes a number of physical processes that are discussed in detail below.

Originally, DI was proposed as a model for the formation of the Solar System \citep{1951PNAS...37....1K}.
It was later superseded by CA as the standard model for planet formation, due to a number of (potential) limitations.
We review these limitations in the following and discuss their relevance.

One of these limitations is very rapid inward migration.
\citet{2011MNRAS.416.1971B} studied the migration of massive planets in gravito-turbulent discs in 2D hydrodynamic simulations, including self-gravity.
They found that these objects migrate to the inner disc on a type~I timescale, without opening a gap.
It was therefore concluded that these planets are unlikely to remain at large separations, at least for the underlying model assumptions (simplified cooling prescription and only a single planet per system).
More recently, \citet{2020MNRAS.496.1598R} performed 3D self-gravitating smoothed particle hydrodynamics (SPH) simulations of single migrating planets in gravitationally unstable discs.
They applied a variable cooling parameter to account for the expected longer cooling time in the inner disc.
The authors found that this effect enables planets to open a gap as soon as they reach the gravitationally stable region in the inner disc; thus, they are able to survive on long timescales.
Both studies, however, neglected the effect of additional companions.
Gravitational interactions between these companions could also facilitate the survival of at least some objects \citep{2021A&A...656A..69E}.

Another related limitation is tidal destruction by Roche lobe overflow \citep[e.g.][]{Kratter2011}.
When a clump moves close to the primary due to migration or it approaches is periapsis on an eccentric orbit, it can be tidally disrupted (see also Sect.~\ref{sec:massloss}).
Whether this occurs depends on the clump evolution (size and mass). 
As a clump contracts (Sect.~\ref{sec:tracks}, its temperature rises.
Once the central temperature exceeds $\approx$~\SI{2000}{K}, the clump  undergoes a dynamical collapse (corresponding to the second collapse in star formation), which makes it stable against tidal disruption \citep{1974Icar...23..319B}.
The pre-collapse timescale depends strongly on the clump's mass, with massive clumps collapsing sooner \citep{helledbodenheimer2011}.
Gas accretion can also shorten this timescale \citep{2012ApJ...756...90V}.
Furthermore, tidal destruction can be avoided if the planet does not move too close to the primary, which can happen if it opens a gap \citep[e.g.][]{2020MNRAS.496.1598R}. Similarly, a clump could become unbound and dissolve again due to irradiation from its environment (i.e. the thermal bath effect, \citealt{2012ApJ...756...90V}).
This fate could be avoided if the clump contracts fast enough.

An additional limitation of DI sometimes brought forward is that it would not be able to form a heavy-element core.
While this is not specifically required, core formation is certainly a possibility  in a clump \citep{Decampli1979,1998ApJ...503..923B,Baehr2019}.
\citet{2008Icar..198..156H} investigated the sedimentation of silicate grains inside isolated clumps of different masses.
They found that protoplanets with masses below \SI{5}{\mj} (Jovian masses) allow for grain sedimentation and core formation, while more massive ones are hotter and in such cases,   core formation via grain settling would not be possible. However, cores in more massive clumps could be a result of fragmentation near spiral arms where dust is accumulated \citep{Baehr2019}. The accretion of heavy elements post formation could further support this process \citep{2006Icar..185...64H}.

Another important challenge in studying DI is the computational cost.
The huge density contrast between the circumstellar disc and the clump's interior is  very challenging numerically when studying fragmenting discs \citep{2010Icar..207..509B,galvagnihayfield2013}.
Advanced hydrodynamical codes are expected to allow for progress in this area in the future \citep[see e.g.][]{Matzkevich2024}.

However, none of these limitations alone are expected to prevent the formation of companions in DI in a general way. Still, at present, these limitations in our understanding of several physical processes occurring during DI are significant. One way to make progress  in to create more realistic model companion formation in the DI scenarios in the future by combining models of the relevant physical processes and including their interplay into a global model. This output can then be confronted with observational results. Differences between these global model predictions and the observations can then be used to constrain the underlying specialized models and eventually improve our understanding. This is our aim in constructing the model presented in this work.

In recent years, DI has again received growing interest as the number of planetary mass companions on wide orbits has increased \citep{Kratter2010,Nielsen2019}.
Other classes of planets that are difficult to explain in the classical CA paradigm have also been identified.
These include giant planets around very low-mass stars, such as GJ~3512~b \citep{Morales2019,2021A&A...645A.139K,2020A&A...633A.116M}, and  massive planets in very young systems, possibly including  AB~Aur~b \citep{2022NatAs...6..751C}. Overall, CA has been studied extensively, also on a population level (\citealt{2004ApJ...604..388I,2021A&A...656A..70E}, see also the reviews in \citealt{2014prpl.conf..691B,2018haex.bookE.143M,Burn2024}). 
Population synthesis projects also exist in the disc instability paradigm \citep{2018MNRAS.474.5036F,2015MNRAS.452.1654N}.
However, these projects are often less mature and/or do not include all the relevant physical processes. 
Disc instability is expected to mainly populate the brown dwarf (BD) and stellar mass regime (e.g. \citealt{Rafikov2005,2010ApJ...708.1585K,Kratter2010,2015MNRAS.454.1940R,Xu2025}).
The exact shape of the population formed in DI is still unknown.
In particular, it is unclear to what extent DI has the capacity to form planetary-mass objects.

A numerical model that includes as many of the relevant physical processes as possible is needed to answer these questions.
In particular, it is important to include the early phase of disc formation in a collapsing molecular cloud core (MCC).
Here, we present a new global model for Disc Instability Population SYnthesis (DIPSY) that includes a number of improvements over previous global models. While being rich in physical mechanism, it is low-dimensional numerically, so that it can still be used to make quantitative statistical predictions that can be compared to observations.
We built on previous works, where an earlier version of the model was used in population syntheses of circumstellar discs \citep{2021A&A...645A..43S,2023A&A...669A..31S}.
The current large model update introduces clumps into the disc if the conditions for fragmentation are satisfied and follows their evolution in detail.

This is the first of a series of papers. In \citet{Schib2025b} (Paper~II), we applied this model to perform a large-scale population synthesis. The current paper is organised as follows.
In Sect.~\ref{sect:discmodel}, we describe the model for the primary star and the circumstellar disc, including the formation stage through infall, the transport of angular momentum, the treatment of self-gravity, the evolution of the central star, and the temperature model.
Section~\ref{sec:clump} details the treatment of fragmentation as well as the evolution of the clumps.
This includes the initial clump mass, mass growth by gas accretion, clump contraction, disc irradiation of the clumps, and mass loss.
In Sect.~\ref{sec:interact}, we discuss the exchange of angular momentum with the disc (leading to orbital migration and eccentricity+inclination damping) as well as gravitational interactions among clumps (N-body).
Section~\ref{sec:sys} demonstrates the capabilities of the model by presenting and discussing several systems of varying complexity.
In Sect.~\ref{sect:Discussion}, we discuss the assumptions and limitations of the model.
We present a summary and our conclusions in Sect.~\ref{sect:Conclusions}.

\section{Disc model}\label{sect:discmodel}
\begin{figure}
  \includegraphics[width=\linewidth]{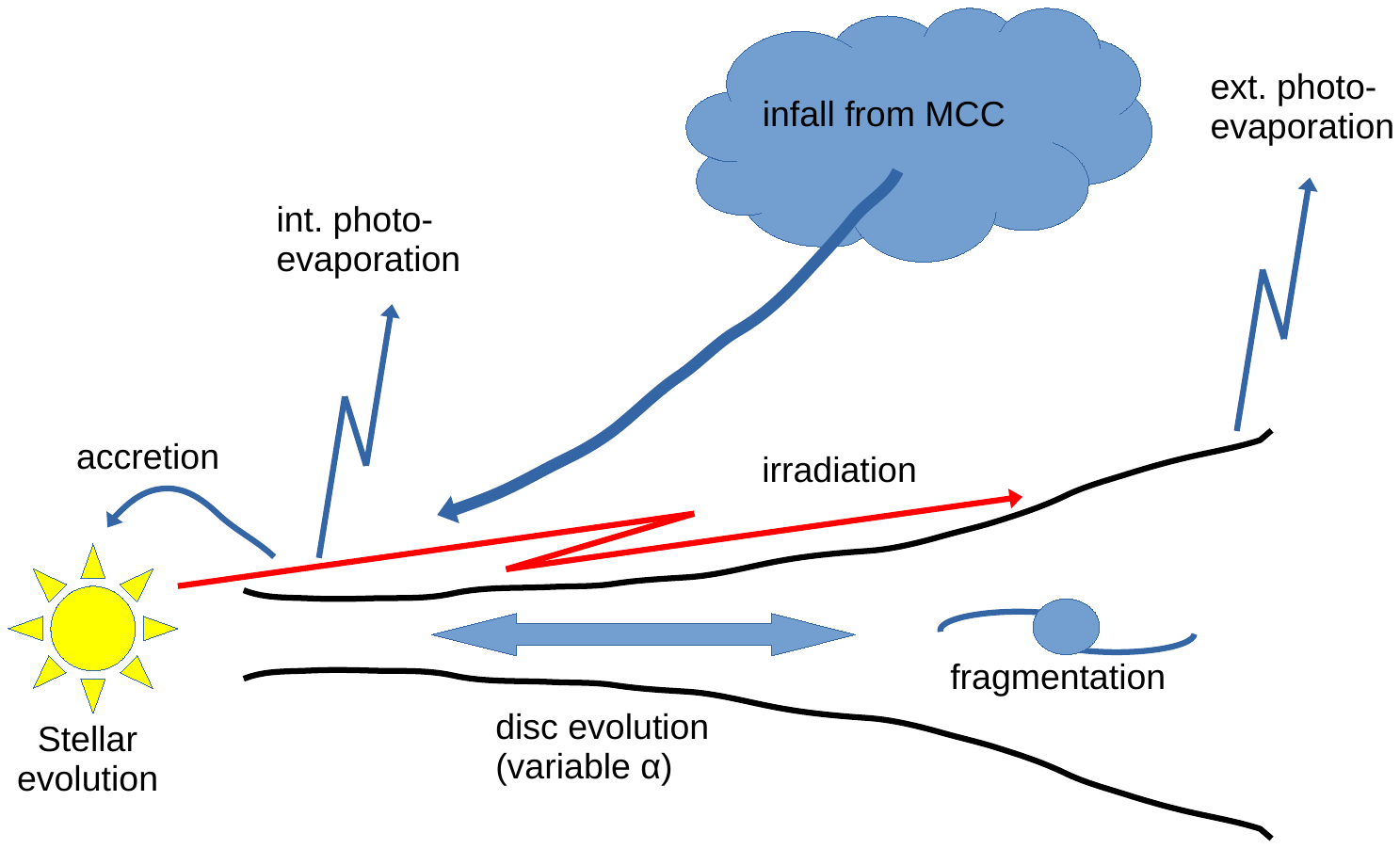}
  \caption{Schematic representation of the star and disc system, including its physical processes.}
  \label{fig:disc}
\end{figure}

The disc model yields a description for the formation and evolution of the circumstellar disc and forms a central part of the overall global model. Figure~\ref{fig:disc} gives a schematic overview of the processes occurring in the disc.
As illustrated in the figure, the model includes the time evolution of the disc and the primary star.
The star and the disc interact via accretion, irradiation, and internal photoevaporation.
The disc further interacts with the environment by infall from the MCC and external photoevaporation.

Adaptations of the disc model used in this work were already employed in previous work: \citet{2021A&A...645A..43S,2022A&A...664A.138S,2023A&A...669A..31S}. We refer to these papers as S21, S22, and S23.
Here, we use a modified and extended version, as described below.  While we focus on the new aspects of the model, for completeness, we describe all the relevant elements of the  model.
We used cylindrical polar coordinates, with $r$ denoting the radial direction (distance from the star). We considered a single primary star of (variable) mass $\mstar$ at the origin. The heliocentric disc's mid-plane is located at $z=\SI{0}{au}$.
The disc is assumed to be rotationally symmetric and is described via the vertically integrated surface density, $\Sigma \equiv \Sigma(r,t)$, which evolves with time $t$ (Sect.~\ref{sec:disc}).
The simulations do not start from a fully formed star-and-disc system. Instead, the star and disc are initialised with a negligible mass and are fed by infalling material from the MCC (Sect.~\ref{sec:infall}). The star is evolved according to pre-calculated stellar evolution tracks (Sect.~\ref{sec:star}).

The disc is assumed to be subject to internal EUV as well as external FUV photoevaporation. Together with accretion, this ultimately leads to the disc's dispersal (Sect.~\ref{sec:disp}).

\subsection{Disc evolution}\label{sec:disc}

The evolution of the disc's surface density $\Sigma$ is calculated by solving the diffusion equation (\citealp{1952ZNatA...7...87L} and \citealp{1974MNRAS.168..603L}):
\begin{equation}\label{eq:evo}
\frac{\partial \Sigma}{\partial t} = \frac{1}{r} \frac{\partial}{\partial r} \left[ 3r^{1/2} \frac{\partial}{\partial r} \left( \nu \Sigma r^{1/2} \right) - \frac{2 \Lambda \Sigma}{\Omega}\right] + S.
\end{equation}
The quantity $\nu$ denotes the kinematic viscosity. We applied the alpha-parametrisation (\citealt{1973A&A....24..337S}, Sect.~\ref{sec:visc}). Then, $\Lambda$ is the torque density distribution used to describe the two-way exchange of angular momentum between the disc and companions (Sect.~\ref{sec:migration}). The angular frequency, $\Omega \equiv \Omega(r,t),$ includes a contribution from the disc's auto-gravitation (Sect.~\ref{sec:auto}). $S \equiv S(r,t)$ stands for a source and sink term:
\begin{equation}\label{eq:sourcesink}
    S(r,t) = S_\mathrm{inf} - S_\mathrm{evap} - S_\mathrm{acc} + S_\mathrm{loss}.
\end{equation}
The contributions to $S$ are: $S_\mathrm{inf}$ the source term for the infalling material from the MCC (Sect.~\ref{sec:infall}), $S_\mathrm{evap}$ the sink term for photoevaporation (Sect.~\ref{sec:disp}), $S_\mathrm{acc}$ the sink term for mass accreted by companions (Sect.~\ref{sec:acc}) and $S_\mathrm{loss}$ the source term for mass returned to the disc by companions undergoing mass loss (Sect.~\ref{sec:massloss}).
Equation~\ref{eq:evo} is solved using the implicit donor cell advection-diffusion scheme with piecewise constant values from \citet{2010A&A...513A..79B}.
We used a grid with \num{2800} logarithmically space grid points extending from \SIrange{3e-2}{3e4}{au}.
At the inner edge, the disc is truncated at \SI{0.05}{au}. The mass flowing across the truncation radius is added to the star, with \SI{10}{\%} considered lost in outflows \citep{2010ApJ...713.1059V}.

\subsection{Viscosity}\label{sec:visc}
The kinematic viscosity $\nu$ is parametrised as \citep{1973A&A....24..337S}:
\begin{equation}
    \nu = \alpha \frac{c_\mathrm{s}^2}{\Omega},
\end{equation}
where $c_\mathrm{s}$ is the isothermal sound speed:
\begin{equation}
    c_\mathrm{s} = \sqrt{\frac{k_\mathrm{B} T}{\mu \mathrm{u}}}.
\end{equation}
Here, $k_\mathrm{B}$ denotes the Boltzmann constant, $T$ is the disc's mid-plane temperature (Sect.~\ref{sec:temperature}), $\mu \equiv 2.3$ is the mean molecular weight and $\mathrm{u}$ is the atomic mass unit.
For the calculation of $\alpha$, we considered different regimes as described below.

\subsubsection{Global instability of the disc}\label{sec:global}
At early times, the disc is supplied by infalling material from the MCC and can become comparable in mass to the star. In this regime, a global gravitational instability can occur \citep[e.g.][]{2011MNRAS.413..423H}. \citet{2010ApJ...708.1585K}\footnote{A more recent study on gravitational instabilities driven by infall is found in \citet{Longarini2025}.} performed a parameter study of rapid accretion in a series of grid-based 3D hydrodynamic simulations. Their results indicated that the disc's behaviour can be characterised by two dimensionless parameters, $\xi$ and $\Gamma$. They parametrised $\alpha$ in the regime in question as follows: 
\begin{equation}
    \alpha_\mathrm{d} = \frac{1}{18(2-k_\Sigma)^2(1+l_j)^2}\frac{\xi^{3/2}}{\Gamma^{1/3}}. 
\end{equation}
We applied this prescription during the infall phase and follow the authors by setting $k_\Sigma = 3/2$ and $l_j = 1$.
The parameters $\xi$ (relating the mass accretion rate of infall to the disc's sound speed) and $\Gamma$ (relating the accretion timescale of the infalling gas to its orbital timescale) are given by: 
\begin{equation}
    \xi = \frac{\dot{M_\mathrm{in}} G}{c_\mathrm{s}^3}~\text{and}~\Gamma = \frac{\dot{M}_\mathrm{in}}{M_\mathrm{*d} \Omega_\mathrm{in}}.
\end{equation}
Here, $\dot{M}_\mathrm{in}$ is the infall rate, $M_\mathrm{*d}$ the combined mass of the star and the disc, and $\Omega_\mathrm{in}$ the angular frequency of the infalling gas at the location where it hits the disc.
The sound speed, $c_\mathrm{s}$, is also evaluated at the same location;
$\xi$ and $\Gamma$ are calculated at each time step in our model.
This makes $\alpha_\mathrm{d}$ constant in $r$ but varying in time.

\subsubsection{Transport of angular momentum by spiral arms}\label{sec:spiral}
Once the infall phase has ended, the prescription from Sect.~\ref{sec:global} is no longer applicable ($\xi=0$).
However, the disc can still be sufficiently massive to be self-gravitating.
In this regime, the transport of angular momentum is governed by spiral arms. We applied the following parametrisation \citep{2010ApJ...713.1143Z,2016MNRAS.461.2257K}:
\begin{equation}
    \alpha_{\mathrm{GI}} = \exp \left( -Q_\mathrm{Toomre}^4 \right),
\end{equation}
by setting $\alpha$ globally and using the minimum of $Q_\mathrm{Toomre}$ (given in Eq.~\ref{eq:qtoomre}) across the disc.
This treatment may somewhat over-predict the transport of angular momentum through spiral arms.
We found that using a radially varying prescription (see also \citealt{2008ApJ...681..375K}) leads to pile-ups of material outside of the unstable regions, which can cause an extreme number of fragmentation event (given that we already observed $\sim~\num{100}$ events in some cases).
Therefore, in this work, we kept the assumption we used in \citet{Schib2021,2023A&A...669A..31S}
However, this is an important topic that needs further investigation.

\subsubsection{Background viscosity}\label{sec:background}
In the absence of global instabilities or spiral arms, we applied a background viscosity, $\alpha_\mathrm{BG}$.
In principle, the model can handle any reasonable value for $\alpha_\mathrm{BG}$.
In the simulations presented in Sect.~\ref{sec:sys} we applied a value of $\alpha_\mathrm{BG} = \num{e-2}$ throughout the disc in order to reproduce observed disc lifetimes and for consistency with \citetalias{2021A&A...645A..43S,2023A&A...669A..31S} and \citetalias{Schib2025b}.
This choice should integrate all mechanisms for the transport of angular momentum, such as effects of disc winds \citep{2013ApJ...769...76B,2014prpl.conf..411T,2016A&A...596A..74S,2023A&A...674A.165W} or hydrodynamic instabilities like the vertical shear instability (see e.g. \citealt{1998MNRAS.294..399U,Klahr2023}).
The value of $\alpha_\mathrm{BG}$ may seem on the higher side compared to values on the order of \num{e-3} seen in the literature \citep{Flaherty2017,Flaherty2018}.
We discuss this topic further in Sect.~\ref{sect:disc_alpha}.
In Appendix.~\ref{app:lv} we study two additional calculations with a lower background viscosity.

In summary, we set $\alpha$ as the maximum of a gravitational $\alpha_\mathrm{G}$ and $\alpha_\mathrm{BG}$,
\begin{equation}\label{eq:alpha}
    \alpha = \mathrm{max}(\alpha_\mathrm{G},\alpha_\mathrm{BG}),
\end{equation}
where $\alpha_\mathrm{G}$ is equal to $\alpha_\mathrm{d}$ during the infall and to $\alpha_\mathrm{GI}$ afterwards.

\subsection{Auto-gravitation}\label{sec:auto}
Since our discs are comparable in mass to their host star at early times, the disc's self-gravity (auto-gravitation) must be considered.
We followed the approach of \citet{2005A&A...442..703H} and assumed that the disc is in vertical hydrostatic equilibrium,
\begin{equation}\label{eq:hydrostat}
    \frac{1}{\rho}\frac{\mathrm{d}p}{\mathrm{d}z} = -\frac{G \mstar}{r^3} z - 4\pi G\Sigma.
\end{equation}
The right-most term is the contribution from the disc's gravity. Then, $\rho \equiv \rho(r,z)$ and $p \equiv p(r,z)$ are the volume density and pressure, respectively.
The disc self-gravity leads to a modification of the angular frequency relative to the Keplerian case,
\begin{equation}\label{eq:angfreq}
    \Omega(r,t)=\bigg[\frac{G \mstar}{r^3}+\frac{1}{r}\frac{\mathrm{d}V_\mathrm{d}}{\mathrm{d}r}\bigg]^{1/2}.
\end{equation}
In Eq.~\ref{eq:angfreq}, $V_\mathrm{d}$ is the gravitational potential of the disc given by \citep{2005A&A...442..703H}:
\begin{equation}
        V_\mathrm{d}(r)=\int_{\rstar}^{\infty}\!-G\frac{4\mathrm{K}\,\big[-\frac{4r/r_1}{(r/r_1-1)^2}\big]}{\left|(r/r_1-1)\right|}\Sigma(r_1)\,\mathrm{d}r_1,
\end{equation}
where $R_*$ is the stellar radius, $\mathrm{K}$ the elliptic integral of the first kind.
The solution of Eq.~\ref{eq:hydrostat} becomes
\begin{equation}
\label{eq:vert}
    \rho(r,z)=\rho_0(r) \exp \bigg(-\bigg(\frac{\left|z\right|}{H_0}+\bigg(\frac{z}{H_1}\bigg)^2\bigg)\bigg),
\end{equation}
with $\rho_0(r) \equiv \rho(r,0)$, where
\begin{equation}
    H_0=\frac{c_\mathrm{s}^2}{4\pi G\Sigma}
\end{equation}
is the contribution of the disc's self-gravity and
\begin{equation}
    H_1=\frac{\sqrt{2}c_\mathrm{s}}{\sqrt{\frac{\mathrm{G} \mstar}{r^3}}}
\end{equation}
is that of the star.
Requiring
\begin{equation}
    \Sigma(r)=\int_{-\infty}^{\infty} \! \rho(r,z) \, \mathrm{d}z.
\end{equation}
Thus, in our convention, we have 
\begin{equation}\label{eq:rhosigma}
    \rho_0(r)=\frac{\Sigma(r)}{\sqrt{2\pi}H},
\end{equation}
which leads to the expression of the vertical pressure scale height,
\begin{equation}
    H=\frac{H_1}{\sqrt{2}}\exp\bigg(\frac{H_1^2}{4H_0^2}\bigg)\bigg(1-\erf\bigg(\frac{H_1}{2H_0}\bigg)\bigg).
\end{equation}
We refer to Sect.~2.2 of \citetalias{2021A&A...645A..43S}, Sect.~3.4.1 of \citet{2005A&A...442..703H} as well as \citet{2000A&A...358..378H} and \citet{1978AcA....28...91P} for additional comments and subtleties of this treatment of auto-gravitation.

The disc's auto-gravitation also has important implications for embedded companions.
They are also subject to the additional acceleration resulting from $V_\mathrm{d}$ and will no longer orbit on Keplerian trajectories (see Sect. \ref{sec:nbody}).
Therefore, it would not make sense to use the usual orbital elements, since these assume Keplerian orbits.
Instead, we usually use the separation, $r$, from the origin to describe a companion's position.
A companion on an eccentric Keplerian orbit (with semi-major axis $a$ and eccentricity $e$) will have a deficit in angular momentum with respect to an equal mass companion with the same semi-major axis on a circular orbit.
The angular momentum is reduced by a factor $\sqrt{1-e^2}$.
We can thus define an `eccentricity' that works in the presence of auto-gravitation by demanding that the angular momentum deficit relative to a circular\footnote{i.e. an orbit where the distance to the primary remain approximately constant} orbit remains the same.

\subsection{Stellar evolution}\label{sec:star}
We applied the stellar evolution tables from \citet{2008ASPC..387..189Y} where the radius and luminosity of an isolated protostar are tabulated. 
We interpolated these tables linearly for a given mass and age.
In addition to the star's intrinsic luminosity, $L_\mathrm{int}$, we included a contribution, $L_\mathrm{acc}$, from accretion of disc material onto the star to calculate the total luminosity,
\begin{equation}\label{eq:ltot}
    L_* = L_\mathrm{int} + L_\mathrm{acc},
\end{equation}
where $L_\mathrm{acc}$ is given by
\begin{equation}\label{eq:lacc}
    L_\mathrm{acc} = f_\mathrm{acc} \frac{G M_* \dot{M}_*}{2 R_*},
\end{equation}
with the stellar accretion rate, $\dot{M}_*$, that is given by the disc model.
The efficiency factor of stellar accretion heating, $f_\mathrm{acc}$, is set to $1/12$ in order to reproduce on average the distribution of observed luminosities of class~0 systems (\citealt{2020ApJ...890..130T},\citetalias{2023A&A...669A..31S}).

\subsection{Infall}\label{sec:infall}
Our simulations were initialised with a seed star and disc. Then the disc was fed by gas infalling from a collapsing MCC. A number of simple prescription how to calculate the source term of infalling material in a spherically symmetric setting exist. Examples are the classical Shu collapse model \citep{1977ApJ...214..488S} or the Bonnor-Ebert sphere model \citep{1955ZA.....37..217E,1956MNRAS.116..351B}.
More complex infall models have been developed on the basis of these ideas, such as the TSC model \citep{1981Icar...48..353C,1984ApJ...286..529T}. While these models provide a relatively simple and intuitive description of the disc formation process, they largely neglect its complex and turbulent nature. Hydrodynamic simulations of star formation show that high angular momentum impacts the disc early, something that is not seen in inside-out collapse. Magnetic fields add another level of complexity that we discuss in the following.

To take into account the chaotic nature of the star formation process, we based our source term on the hydrodynamic population synthesis study of discs by \citet{2018MNRAS.475.5618B}. The paper analyses the evolution and properties of discs formed in a 3D hydrodynamic star cluster simulation. Stellar masses, disc masses, and disc radii are given as a function of time for 183 synthetic protostars.
In \citetalias{2021A&A...645A..43S} we discuss in detail how we select a sample of these systems in order to construct probability distributions for initial stellar mass, disc mass,  infall rate and disc size. The first three of these quantities are considered correlated: we used multi-variate distributions to obtain the initial values for these quantities.
This data was then used as input quantity to perform a population synthesis of forming discs and analyse their properties. As found in \citetalias{2021A&A...645A..43S} this baseline run `hydro' leads to massive ($\sim \SI{0.3}{\msun}$) and large ($\sim \SI{200}{au})$ discs. Almost half of these discs were found to fragment.
However, magnetic fields, which were not considered in \citet{2018MNRAS.475.5618B}, are thought to alter the disc formation process. Discs are found to be smaller in magnetohydrodynamic (MHD) collapse simulations than in pure hydrodynamic simulations, though the specifics depend on which aspects of magnetic fields are considered.
In order to investigate this effect, another run  based on MHD was performed in \citetalias{2021A&A...645A..43S}. There, we imposed much smaller infall radii (the location where the infalling material hits the disc) in such a way that the distribution of early disc sizes agrees with the prediction of an MHD simulation of disc formation \citep{Hennebelle2016}. The discs formed in this simulation are less massive ($\sim \SI{0.1}{\msun}$), much smaller ($~\sim \SI{40}{au}$) and none of them fragment.
Interestingly, the sizes of observed Class~0 discs \citep{2020ApJ...890..130T} lies between the early disc size in `hydro' and `MHD' runs.
This led us to perform an additional set of simulations with an additional modification of the infall radii set to fit the observed Class~0 disc sizes.
Together with the choice of $f_\mathrm{acc}$ (Eq.~\ref{eq:lacc}), this leads to the run `OBS\_REDIRR' presented in \citetalias{2023A&A...669A..31S}, which exhibits a distribution of both early disc radii and luminosities in agreement with observations. In other words, in this work we use infall radii that lead to discs with sizes that agree with the observations of \citet{2020ApJ...890..130T}. The infall radii found in this way are discussed in \citetalias{Schib2025b}.
The source term for the infall $S_\mathrm{inf}$ is given by (Eq.~32 of \citetalias{2021A&A...645A..43S}):
\begin{equation}
    S_\mathrm{inf} (r,t) = \frac{\dot{M}_\mathrm{in}}{2 \sqrt{2} \pi^{3/2} R_\mathrm{i} \sigma_\mathrm{i}} \exp{\left[ - \left( \frac{r - R_\mathrm{i}(t)}{\sqrt{2} \sigma_\mathrm{i}} \right) \right]}.
\end{equation}
Here, $R_\mathrm{i}$ is the infall location and the width $\sigma_\mathrm{i}$ is chosen as $R_\mathrm{i}/3$.

\subsection{Temperature model}\label{sec:temperature}
Our temperature model is based on the vertical energy conservation of the disc  which is in hydrostatic equilibrium (Eq. \ref{eq:hydrostat}).
We considered the effects of the following processes: viscous heating, irradiation from the central star (including an accretion term, Eq.~\ref{eq:ltot}), shock heating from the gas falling from the MCC onto the disc and a constant background irradiation from the environment.
We follow \citet{1994ApJ...421..640N,2005A&A...442..703H} and consider both optically thick and an optically thin regimes and assume an energy balance at the disc's surface,
\begin{equation}\label{eq:tsurf}
        \sigma T_\mathrm{S}^4 = \frac{1}{2}\left( 1 + \frac{1}{2 \Sigma \kappa_\mathrm{p}} \right) (\dot{E}_\nu + \dot{E}_\mathrm{s}) + \sigma T_\mathrm{env}^4 + \sigma T_\mathrm{irr}^4,
\end{equation}
where $T_\mathrm{S}$ is the disc's surface temperature.
This leads to an expression for the disc's mid-plane temperature:
\begin{equation}\label{eq:tmid}
    \begin{aligned}
        \sigma T_\mathrm{mid}^4 &=
        \frac{1}{2} \left[ \left( \frac{3 \Sigma \kappa_\mathrm{R}}{8} + \frac{1}{2 \Sigma \kappa_\mathrm{P}} \right) \dot{E}_\nu + \left( 1 + \frac{1}{2 \Sigma \kappa_\mathrm{P}} \right) \dot{E}_\mathrm{s} \right] \\
        &+ \sigma T_\mathrm{env}^4 \\
        &+ \sigma T_\mathrm{irr}^4.
    \end{aligned}
\end{equation}
In Eq.~\ref{eq:tmid}, $\kappa_\mathrm{R} \equiv \kappa_\mathrm{R} (\rho_0,T_\mathrm{mid})$ and $\kappa_\mathrm{P} \equiv \kappa_\mathrm{P} (\rho_0,T_\mathrm{mid})$ are the Rosseland and Planck mean opacities. Then, $\kappa_\mathrm{P}$ and $\kappa_\mathrm{R}$ are calculated based on \citet{2014A&A...568A..91M} for the gas and \citet{2003A&A...410..611S} for the dust. The calculation is done analogous to \citet{2017ApJ...836..221M,2019ApJ...881..144M}, assuming `normal silicate' dust grains made of homogeneous spheres (per the NRM model in \citealt{2003A&A...410..611S}).
We assume a dust-to-gas ratio of \num{0.01}.
In Eq.~\ref{eq:tmid}, $\dot{E}_\nu = \Sigma \nu (T_\mathrm{mid}) \left( r \frac{\mathrm{d} \Omega}{\mathrm{d} r} \right)^2$ is the viscous heating term, $\dot{E}_\mathrm{s} = S_\mathrm{inf} (r \Omega)^2/2$ the shock heating term due to infall \citep{2016MNRAS.461.2257K}.
The temperature contribution due to stellar irradiation, $T_\mathrm{irr}$, is calculated as in \citet{2005A&A...442..703H}:
\begin{equation}\label{eq:irr}
    T_\mathrm{irr} = T_* \left[ \frac{2}{3 \pi} \left( \frac{R_*}{r} \right)^3 + \frac{1}{2} \left( \frac{R_*}{r} \right)^2 \left(\frac{\mathrm{d} \ln(H)}{\mathrm{d} ln(r)}-1 \right) \right]^{1/4},
\end{equation}
where the effective stellar temperature $T_*$ is calculated as
\begin{equation}
    T_* = \left( \frac{L_*}{4 \pi R_*^2 \sigma} \right)^{1/4}, 
\end{equation}
with $L_*$ from Eq.~\ref{eq:ltot}.
In Eq.~\ref{eq:irr}, we set $\mathrm{d} \ln(H)/\mathrm{d} \ln(r) \equiv 9/7$ \citep{chianggoldreich1997,Fouchet2012}.

\subsection{Photoevaporation and disc dispersal}\label{sec:disp}

We assumed that the discs are subject to external and internal photoevaporation.
Our model of external photoevaporation is based on \citet{2003ApJ...582..893M}, which considers FUV irradiation by  massive stars.
We applied the following sink term,
\begin{equation}
    S_\mathrm{ext}(r,t) =
    \begin{cases}
        S_\mathrm{wind} \left( 1 - \frac{1}{1+sm_\mathrm{ext}^{20}} \right),&\text{if } r > 0.1 \beta_\mathrm{M} r_\mathrm{mI}\\
        0, & \text{otherwise},
    \end{cases}
\end{equation}
with a smoothing term $sm_\mathrm{ext} = \frac{r}{\beta_\mathrm{M} r_\mathrm{g,ext}}$ and $r_\mathrm{g,ext}(t) = G \mstar(t) / c_\mathrm{s,ext}^2$ the gravitational radius.
We set $c_\mathrm{s,ext} = \SI{2.5}{km.s^{-1}}$, $\beta_\mathrm{M} = 0.14$ and $S_\mathrm{wind} = \SI{2.8e-8}{g.cm^{-1}.yr^{-1}}$, giving an evaporation rate of $\SI{e-8}{\msun.yr^{-1}}$ for a disc that extends to $\SI{1000}{au}$.
This value is relatively low, though comparable with previous work \citep{2003MNRAS.342.1139A,2009A&A...501.1139M}.

For the internal photoevaporation, we closely follow \citet{2001MNRAS.328..485C}, an EUV model based on the `weak stellar wind' case studied in \citet{1994ApJ...428..654H}. We used a sink term given by 
\begin{equation}
    S_\mathrm{int}(r,t) = 2 c_\mathrm{s,int} d_\mathrm{r} \mathrm{u} sm_\mathrm{int},
\end{equation}
with a sound speed $c_\mathrm{s,int} = \SI{11.1}{km.s^{-1}}$,
\begin{equation*}
    d_\mathrm{r} = \num{1.8e4} (\mstar(t)/\msun)^{-0.25} \left(\frac{r_\mathrm{g,int}}{\num{e14}}\right)^{-1.5} \left(\frac{r}{r_\mathrm{g,int}}\right)^{-2.5}.
\end{equation*}
$sm_\mathrm{int}$ is a smoothing factor of
\begin{equation}
    sm_\mathrm{int} =
    \begin{cases}
        1 - \left( 1 + \left(\frac{r}{0.14 r_\mathrm{g,int}}\right)^{20} \right) ^{-1}, & \text{if } r > r_\mathrm{int}\\
        0, & \text{otherwise},
    \end{cases}
\end{equation}
where $r_\mathrm{int} = 0.07 r_\mathrm{g,int}, r_\mathrm{g,int}(t) = G \mstar(t) / c_\mathrm{s,int}^2$.

\section{Fragmentation and clump evolution}\label{sec:clump}
\begin{figure}
  \includegraphics[width=\linewidth]{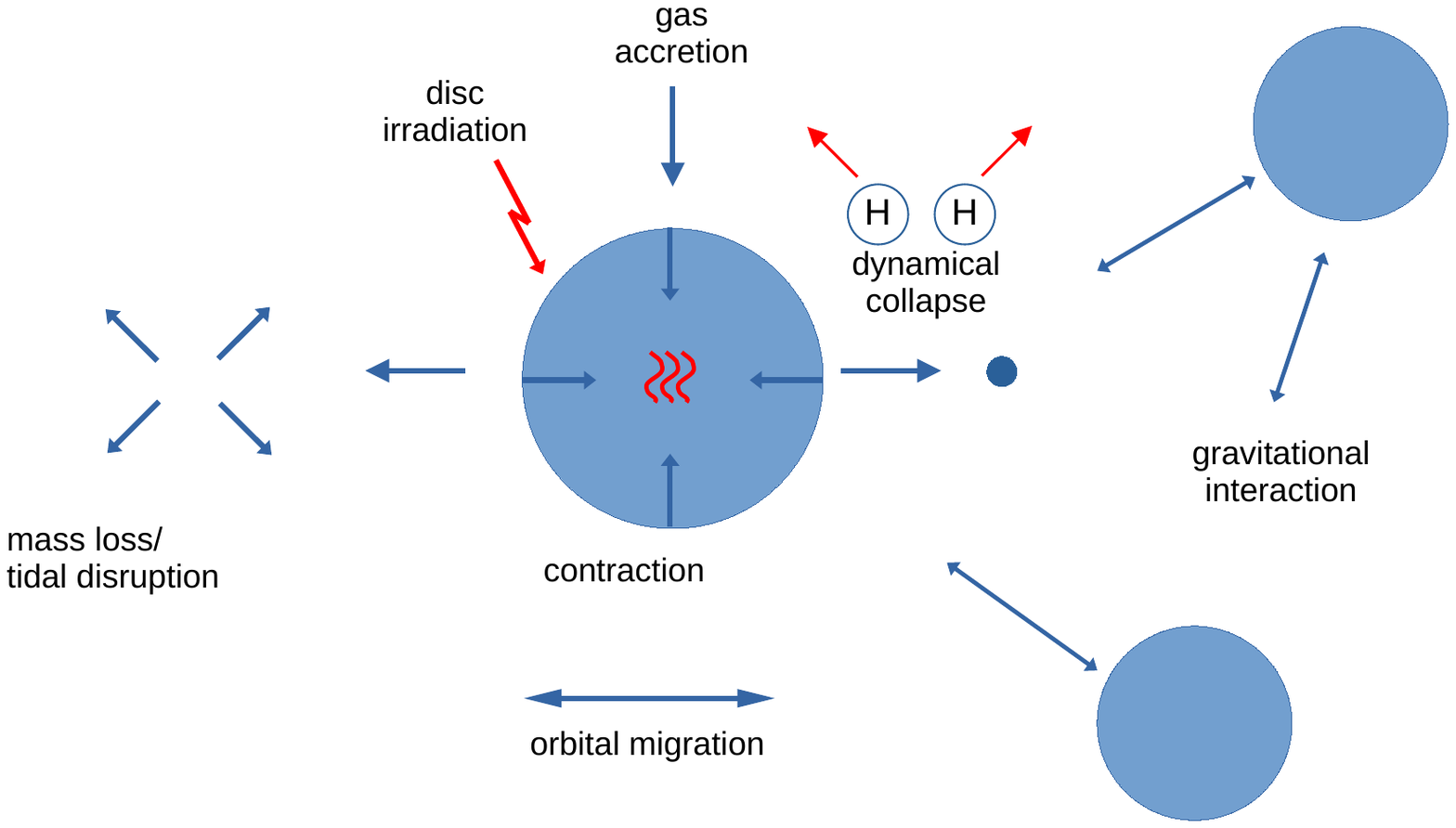}
  \caption{Schematic overview of the physical processes related to fragmentation and the evolution of companions that are included in the global model.}
  \label{fig:clumpevo}
\end{figure}

In this section, we discuss the topic of fragmentation, and how we handled it in the model.
This includes the formation of a clump and its subsequent evolution.
Figure~\ref{fig:clumpevo} gives a schematic overview of the relevant processes.
When the disc fragments, one or several clumps are inserted.
The clumps can accrete gas (Sect.~\ref{sec:acc}) and evolve in their internal structure according to pre-calculated clump evolution tracks (Sect.~\ref{sec:tracks}) under the influence of disc irradiation (Sect.~\ref{sec:irrad}).
Mass loss of clumps is discussed in Sect.~\ref{sec:massloss}.
Clumps can undergo orbital migration (Sect.~\ref{sec:migration}) and interact gravitationally with each other (Sect.~\ref{sec:nbody}).
\subsection{Fragmentation}\label{sect:frgmentation}
Fragmentation is the breaking up, due to self-gravity, of parts of the disc into bound objects which we call clumps or fragments.
The degree to which a disc is self-gravitating can be quantified by the $Q_\mathrm{Toomre}$ parameter \citep{1964ApJ...139.1217T}:
\begin{equation}\label{eq:qtoomre}
    Q_\mathrm{Toomre} = \frac{c_\mathrm{s}\kappa}{\pi G \Sigma}. 
\end{equation}
It results from the stability analysis of a thin, differentially rotating fluid sheet. A $Q_\mathrm{Toomre}< Q_\mathrm{crit} \sim \num{1}$ permits exponentially growing modes.
In Eq.~\ref{eq:qtoomre}, $\kappa$ denotes the epicyclic frequency:
\begin{equation}\label{eq:epi}
    \kappa = \sqrt{4 \Omega^2 + 2 r \Omega \frac{\mathrm{d} \Omega}{\mathrm{d}r}}, 
\end{equation}
equal to $\Omega$ only in Keplerian discs.
For fragmentation, $Q_\mathrm{Toomre} < Q_\mathrm{crit}$ is not sufficient. Such discs can remain in a state of marginal stability, where the transport of angular momentum is dominated by spiral arms \citep{1971ApJ...168..343H,1988A&A...195..105B,2004MNRAS.351..630L,Cossins2009,Forgan2011a,Rice2016}.
A disc fragments only if saturation mechanisms fail to prevent the growth of perturbations.
There is a wealth of literature on fragmentation, including \citet{1965MNRAS.130...97G,2001ApJ...553..174G,2005ApJ...628..817M,Lodato2005,2006MNRAS.373.1563K}
An extensive review can be found in \citet{2016ARA&A..54..271K}.
In our context, it is important to consider two regimes of fragmentation.

\subsubsection{Infall-dominated regime of fragmentation}
During the infall phase, material from the MCC can reach the disc at a rate greater than it can be transported away. If this occurs, the disc invariably fragments  \citep{2009ApJ...695L..53B}. We call this the `infall-dominated' regime.
The $\xi$ and $\Gamma$ parameters introduced in Sect.~\ref{sec:global} can be employed to check if we are in this regime. The condition is given in \citet{2010ApJ...708.1585K} (see also Sect.~3.6 of \citealt{2016ARA&A..54..271K}):
\begin{equation}\label{eq:infdom}
    \Gamma < \frac{\xi^{2.5}}{850}.
\end{equation}
In our model, we assume the disc fragments if $Q_\mathrm{Toomre} < 1$ in the infall-dominated regime.

\subsubsection{Cooling-dominated regime of fragmentation}
If the condition from Eq.~\ref{eq:infdom} is not satisfied (in particular after the infall phase), the disc fragments  only if it can cool efficiently.
\citet{2001ApJ...553..174G} performed shearing sheet simulations of a razor thin disc to demonstrate that this is the case, if the cooling timescale is less than a few times the orbital timescale, following the expression
\begin{equation}\label{eq:gammie}
    t_\mathrm{cool} \lesssim \beta_\mathrm{c} \Omega^{-1},
\end{equation}
a condition known as the Gammie criterion. $\beta_\mathrm{c} \approx 3$ is the critical cooling parameter as determined by \citet{2017ApJ...847...43D}, see also \citet{2017ApJ...848...40B}.
We employ the following cooling timescale  \citep{2012A&A...547A.112M}:
\begin{equation}\label{eq:tcool}
    t_\mathrm{cool} = \frac{3 \gamma \Sigma c_\mathrm{s}^2}{32 (\gamma - 1) \sigma T^4}\tau_\mathrm{eff},
\end{equation}
where $\tau_\mathrm{eff} = \kappa_\mathrm{R} \Sigma /2 + 2/(\kappa_\mathrm{R} \Sigma) $ is an effective optical depth, $\gamma \equiv 1.45$ is the adiabatic index.
We therefore assume that the disc fragments if $Q_\mathrm{Toomre} < 1$ and Eq.~\ref{eq:gammie} are satisfied simultaneously.

In summary, our condition for fragmentation is as follows: $Q_\mathrm{Toomre}<1$ and either Eq.~\ref{eq:infdom} or \ref{eq:gammie} must be satisfied. These conditions are checked throughout the disc at every time step.

\subsection{Initial fragment mass and insertion of fragments}\label{sect:initialfragmass}
If the disc fragments, the initial fragment mass $M_\mathrm{F}$ (given in Eq.~\ref{eq:mfraginit} below) is removed from the disc at the location (annulus), where $Q_\mathrm{Toomre}$ is minimal at the time of fragmentation. (For population synthesis calculations, the procedure is slightly different, as explained in Sect.~4.4 of \citetalias{Schib2025b}).
The clump's properties (aside its mass and age) are determined by means of evolution tracks, as described in Sect.~\ref{sec:tracks}.
A clump of the same mass is inserted at this location $r_\mathrm{c}$.
In order to avoid unphysical effects by removing mass from the disc instantly, we instead removed the initial fragment mass on a fiftieth of an orbital timescale. We found that varying this value over reasonable ranges does not strongly affect the outcomes.
We apply the following sink term in Eq.~\ref{eq:sourcesink}:
\begin{equation}
    S_\mathrm{acc}(r) = \frac{\dot{M}_\mathrm{insert}}{\sqrt{2 \pi \delta}} \exp{\left(-\frac{(r-r_\mathrm{c})^2}{2 \delta}\right)},
\end{equation}
with $\dot{M}_\mathrm{insert} = 50 M_\mathrm{F} \Omega(r_\mathrm{c})/(2 \pi )$.
For $\sqrt\delta$, we chose $2 \sqrt[3]{M_\mathrm{F}/3\mstar}$ (the exact value of this width is not very important).

How massive emerging clumps are is important, since the pre-collapse timescale has a strong mass dependency (Sect.~\ref{sec:massloss}).
A rough estimate for the initial fragment mass is the Toomre mass \citep{Nelson2006}:
\begin{equation}
    M_\mathrm{Toomre} = \frac{\pi c_\mathrm{s}^4}{G^2 \Sigma}.
\end{equation}
\citet{2011MNRAS.417.1928F} performed more detailed calculations and derived a measure for the local Jeans mass inside the spiral structure of a self-gravitating disc. In our conventions (note the comments in Sect.~2.9.2 in \citetalias{2021A&A...645A..43S}) it is equivalent to
\begin{equation}\label{eq:mjeans}
    M_\mathrm{J,FR} = \frac{4}{3} \frac{2^{1/4} \pi^{11/4} c_\mathrm{s}^3 H^{1/2}}{G^{3/2} \Sigma^{1/2} \sqrt{1 + \beta_\mathrm{cool}^{-1}}},
\end{equation}
where $\beta_\mathrm{cool} = t_\mathrm{cool} \Omega$.
The cooling timescale $t_\mathrm{cool}$ is given in Eq.~\ref{eq:tcool}.
A different estimate of the initial fragment mass is given in \citet{2010Icar..207..509B}. The authors performed a calculation that considers the density perturbation near the corotation of a spiral arm.
They initial clump mass in the context of a fragmenting spiral arm was found to be 
\begin{equation}\label{eq:mfraginit}
    M_\mathrm{F} = \frac{1.6 c_\mathrm{s}^3}{G \Omega}.
\end{equation}
We observe that there is a substantial difference between these three estimates, with $M_\mathrm{F}$ being by far the lowest.
For a Keplerian disc with $Q_\mathrm{Toomre}=1$, we have
\begin{equation*}
    M_\mathrm{F} \approx 0.2 M_\mathrm{J}^\mathrm{Sphere} \approx 0.16 M_\mathrm{Toomre} \approx 0.04 M_\mathrm{J,FR},
\end{equation*}
where $M_\mathrm{J}^\mathrm{Sphere}$ is the Jeans mass given in Eq.~13 of \citet{Nelson2006}.
\citet{2009ApJ...695L..53B} performed 3D SPH simulations with radiative cooling and found good agreement of their analytic expression ($M_\mathrm{F}$)  with the simulation results.
This has also been investigated by \citet{Tamburello2015}, albeit in a different context. These authors performed simulations of isolated galaxies using an N-body + SPH code and also found initial fragment masses comparable to or even lower than $M_\mathrm{F}$.
A possible interpretation of the much higher value of $M_\mathrm{J, FR}$ is that it does not represent the mass of a clump at the time of fragmentation, but instead, after a few orbits, where it could have accreted a substantial amount of gas.
When including magnetic fields, we note that the gravitational instability dynamo may lead to even lower fragment masses \citep{2021NatAs...5..440D,2023MNRAS.525.2731K}.
Given these results, we use $M_\mathrm{F}$ as the nominal initial mass for our clumps.
The influence of a higher initial fragment mass is investigated in \citetalias{Schib2025b}.

During the short period where the fragment is set up, it is kept at a constant separation and is not allowed to interact with the disc (except for the mass removal). No other fragments are allowed to form inside of a Hill radius (for $M_\mathrm{F}$) during this time.

\subsection{Gas accretion}\label{sec:acc}

We applied the model for gas accretion as presented in \citetalias{2022A&A...664A.138S}. This model is based on the Bondi and Hill accretion regimes investigated in \citet{2008ApJ...685..560D} (DL08). These authors performed 3D nested grid hydrodynamic simulations of migrating planet undergoing rapid gas accretion. They estimated the accretion rate onto the companion in their Eq.~9, expressed as
\begin{equation}\label{eq:macc}
    \dot{M}_\mathrm{c} \sim \frac{\Sigma}{H}\Omega R_\mathrm{f}^3.
\end{equation}

Here, $R_\mathrm{f}$ is the radius of the feeding zone of a companion of mass $M_\mathrm{c}$ at a separation $r_\mathrm{c}$.
It is taken to be the smaller of either the Bondi radius $R_\mathrm{B} = G M_\mathrm{c}/c_\mathrm{s}^2$ or the Hill radius\footnote{In practice, we used a more accurate expression for the Hill radius if $M_\mathrm{c}/M_* > 0.01$. This is discussed in Appendix~\ref{app:rhill}.} $R_\mathrm{H} = r_\mathrm{c} \sqrt[3]{ M_\mathrm{c}/\left(3 \mstar\right)}$.
The accretion rates in the Bondi and Hill regime then become:
\begin{equation}\label{eq:mb}
    \dot{M}_\mathrm{B} = C_\mathrm{B} \Omega \frac{\Sigma r_\mathrm{c}^2}{\mstar}\left(\frac{r_\mathrm{c}}{H}\right)^7\left(\frac{M_\mathrm{c}}{\mstar}\right)^2 M_\mathrm{c},
\end{equation}
\begin{equation}\label{eq:mh}
    \dot{M}_\mathrm{H} = \frac{1}{3}C_\mathrm{H}\Omega \frac{\Sigma r_\mathrm{c}^2}{\mstar}\left(\frac{r_\mathrm{c}}{H}\right),
\end{equation}
where $C_\mathrm{B}$ and $C_\mathrm{H}$ are dimensionless coefficients of order unity.
\citepalias{2008ApJ...685..560D} found that the accretion rate onto the protoplanet agrees well with $\min (\dot{M}_\mathrm{B},\dot{M}_\mathrm{H})$ as long as the disc is able to supply enough gas.
Once the local reservoir is depleted, the accretion rate drops.

In \citetalias{2022A&A...664A.138S} we derived an accretion model based on Eq.~\ref{eq:macc}.
Instead of using global values of $\Sigma$ and $\Omega$, we calculate the contributions from each grid cell inside the feeding zone separately.
The gas is then removed self-consistently from the disc at the location from where it was accreted.
We obtain the following accretion rate (Eq.~D.5 in \citetalias{2022A&A...664A.138S}):
\begin{equation}\label{eq:mdot}
\begin{aligned}
    \dot{M}_\mathrm{B,H}
    &= \sqrt{\pi} C_\mathrm{B,H} \int_{r_\mathrm{c}-R_\mathrm{f}}^{r_\mathrm{c}+R_\mathrm{f}}\!\rho_0(r) H_1 \exp{\left(\frac{H_1^2}{4 H_0^2}\right)} \times\\
    &\left(\erf{\left[\frac{\sqrt{R_\mathrm{f}^2-(r-r_\mathrm{c})^2}}{H_1} + \frac{H_1}{2 H_0}\right]}-\erf{\left[\frac{H_1}{2 H_0}\right]}\right)\,v_\mathrm{rel}\,\mathrm{d}r.
\end{aligned}
\end{equation}
In Eq.~\ref{eq:mdot}, $C_\mathrm{B,H}$ denotes the numerical factor, calibrated with the results from \citetalias{2008ApJ...685..560D} to be $C_\mathrm{B}=10$ and $C_\mathrm{H}=0.19$, respectively in \citetalias{2022A&A...664A.138S}.
The quantities $H_0$ and $H_1$ are related to the disc's auto-gravitation and are described in Sect.~\ref{sec:auto}.
$v_\mathrm{rel} = \left|r \Omega - r_\mathrm{c} \Omega_\mathrm{c}\right|$ is the velocity of the gas relative to the companion.

In \citetalias{2022A&A...664A.138S} we compare the accretion model, together with our migration model with the results from two different hydrodynamic simulations of accreting, migrating companions. We typically find a reasonable agreement for a range of parameters. 

\subsection{Clump evolution tracks}\label{sec:tracks}
\begin{figure*}
    \begin{subfigure}[pt]{0.49\textwidth}
    \includegraphics[width=\linewidth]{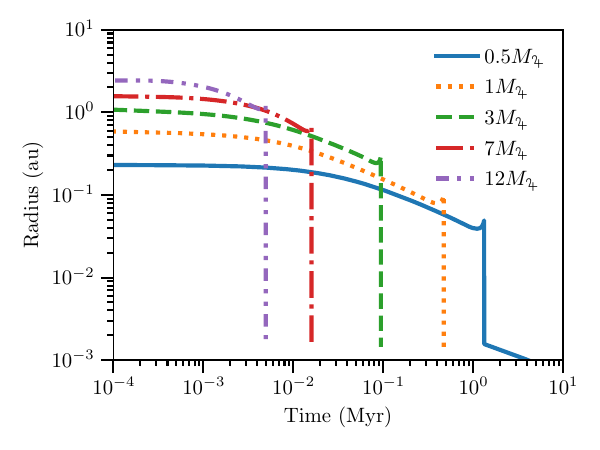}
    \end{subfigure}
    \begin{subfigure}[pt]{0.49\textwidth}
    \includegraphics[width=\linewidth]{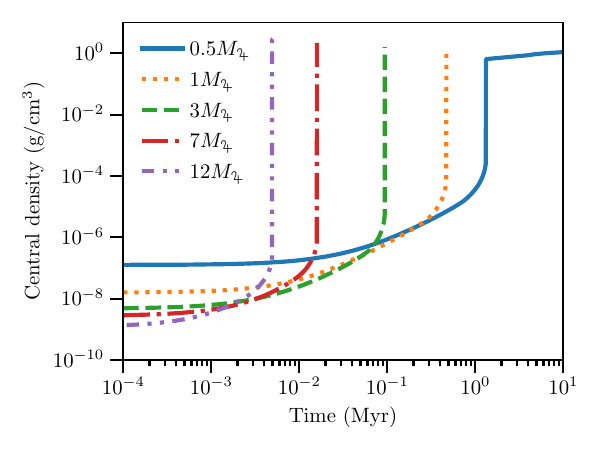}
    \end{subfigure}
    \begin{subfigure}[pt]{0.49\textwidth}
    \includegraphics[width=\linewidth]{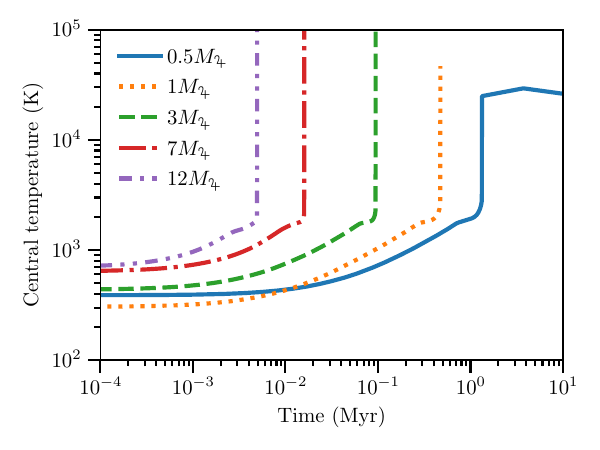}
    \end{subfigure}
    \begin{subfigure}[pt]{0.49\textwidth}
    \includegraphics[width=\linewidth]{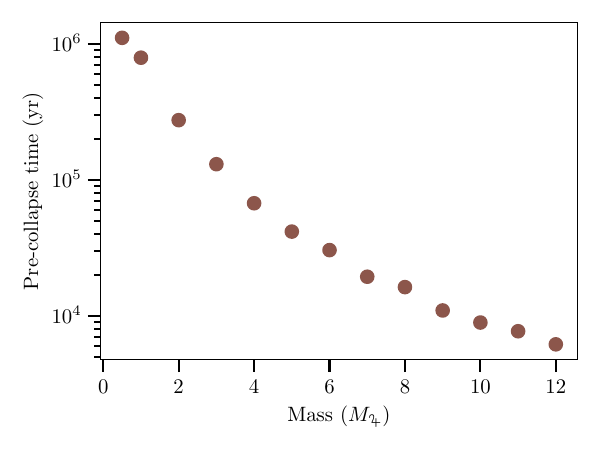}
    \end{subfigure}
    \caption{Evolution tracks of isolated clumps.
             Top left: Clump radius.
             Top right: Central density.
             Bottom left: Central temperature.
             Bottom right: Pre-collapse time for all clump masses.
             These tracks were first published in \citet{2019MNRAS.488.4873H}.}
    \label{fig:clump}
  \end{figure*}

Just after formation, clumps are extended, \SI{}{au}-sized objects of low density. They will contract and heat up on a timescale that strongly depends on their mass.
During the time that a clump is young and extended, it is prone to tidal destruction. However, if the interior reaches temperatures of $\sim \SI{2000}{K}$, the clump undergoes a dynamical collapse due to the dissociation of molecular hydrogen \citep{1974Icar...23..319B}. The radius then shrinks by about  three orders of magnitude to a few Jovian radii ($\rj$).
In order to account for this evolution, we employed interior evolution tracks of isolated clumps.
These tracks are the result of 1D calculations of clumps published in \citet{2019MNRAS.488.4873H}.
The code used solves the standard stellar structure equations on a fully implicit, adaptive grid \citep{2006Icar..185...64H,2008Icar..195..863H,2012ApJ...756...90V,vazankovetz2013,2015ApJ...803...32V}.
It assumes a protosolar hydrogen-helium composition and the \citet{1995ApJS...99..713S} equation of state.
Heat transport inside the clump is modelled by convection, conduction or radiation, depending on the local conditions.
In the radiative regions, opacities from \citet{1985Icar...64..471P} are used.
The clumps evolve in isolation (i.e. with a zero pressure boundary condition), which means that no exchange of mass or irradiation are considered.
Tracks are available for a number of masses, ranging from \SIrange{0.5}{12}{\mj}.
Since the clump masses in our simulations evolve with time,  we linearly interpolate the quantities from the evolution tracks.
This allows us to determine the clumps' (and later compact objects') properties, such as radius and temperature, at any time during their evolution.
Figure~\ref{fig:clump} shows the time evolution of the radius, central density and central temperature for a selection of clump masses. Also shown is the pre-collapse time for all masses. We define the pre-collapse time as the age of the clump when its central temperature reaches \SI{2000}{K}.
At this time, the clump radius plummets, while the central density and temperature increase.

\subsection{Clump irradiation}\label{sec:irrad}
Our evolution tracks were calculated for isolated clumps. However, when clumps form, they are embedded in the surrounding circumstellar disc. The clumps are therefore irradiated by the surrounding disc material. This influences their evolution, and this effect is not considered in the tracks.

In the absence of tabulated tracks including irradiation we follow an approximate approach inspired by  \citet{2019MNRAS.488.4873H}\footnote{We note that there are some typos in their Sect.~2.7.}. The authors assume that, because of the irradiation, the clump's evolution is slowed down: The clump's `internal time', $\tau$, passes more slowly than the physical time $t$ depending on the incident radiation. For the clump's total energy $E$ we have:
\begin{equation}\label{eq:irr_iso}
    \frac{\mathrm{d}E(\tau)}{\mathrm{d}\tau} = - L_0(\tau),
\end{equation}
where $L_0(\tau)$ is the clumps radiative luminosity in isolation.
If the clump is embedded in the disc, its energy balance is modified by the incident irradiation from the disc, and at time $t$ it must hold:
\begin{equation}\label{eq:irr_disc}
    \frac{\mathrm{d}E(t)}{\mathrm{d}t} = - L(t) = - L_0(t) + 4 \pi R_\mathrm{c}^2(t)\sigma_\mathrm{B} T_\mathrm{mid}^4(t),
\end{equation}
where the right-most term denotes the radiation incident on the clump from the disc ($R_\mathrm{c}$ is the clump's radius and $T_\mathrm{mid}$ is the disc's mid-plane temperature evaluated at the clump's location).
The clump's radius and luminosity for a given clump mass and age are linearly interpolated from the evolution tables (Sect.~\ref{sec:tracks}).
Equation~\ref{eq:irr_disc} expresses that, in order to infer the clump's effective luminosity $L$, its luminosity in isolation must be reduced by the incident luminosity from the disc.
In order to find the relationship between $t$ and $\tau$, we divide Eq.~\ref{eq:irr_disc} by Eq.~\ref{eq:irr_iso} and obtain
\begin{equation}\label{eq:tauirr}
    \frac{\mathrm{d}E(t)}{\mathrm{d}t} \left( \frac{\mathrm{d}E(\tau)}{\mathrm{d}\tau} \right)^{-1} = \frac{- L(t)}{-L_0(\tau)} \\ = \frac{\mathrm{d}\tau}{\mathrm{d}t}
    = \frac{L_0(t) - 4 \pi R_\mathrm{c}^2(t) \sigma_\mathrm{B} T_\mathrm{mid}^4(t)}{L_0(\tau)}.
\end{equation}
In the second equality we required $\mathrm{d}E(t) = \mathrm{d}E(\tau)$ since  we want to evaluate the evolution tables at a time where the change in total energy of the isolated case is the same as in the irradiated case.
Equation~\ref{eq:tauirr} allows us to infer $\tau$ for a given time $t$. For example for $\tau_0 = t_0 = \SI{0}{s}$, at $t_1$, we get
\begin{equation}
    \tau_1 = \tau_0 + \frac{\mathrm{d}\tau}{\mathrm{d}t} (t_1 - t_0) = \frac{\mathrm{d}\tau}{\mathrm{d}t} t_1.
\end{equation}
Knowing the relationship between $t$ and $\tau$ allows us to calculate the properties of the irradiated clump: for a clump with internal age $\tau_1$ the evolution tables are evaluated at $\frac{\mathrm{d}\tau}{\mathrm{d}t} t_1$.

This approach is a reasonable approximation when  $T_\mathrm{mid}$ is much lower than the clump's effective temperature $T_\mathrm{eff}$; however, it becomes problematic if the two reach comparable values.
For example, if they are the same, Eq.~\ref{eq:tauirr} gives $\tau = const$.
In other words, the clump stops evolving.
In a realistic scenario, we would expect the clump to adapt to a different interior structure, possibly with a larger radius or losing some of its mass.
The situation becomes even more problematic if $T_\mathrm{mid}$ exceeds $T_\mathrm{eff}$.
In practice, we kept $\tau$ constant in this case, and instead compared $T_\mathrm{mid}$ with the clump's central temperature.
If the disc's temperature exceeds even that, we assumed that the clump is thermally destroyed \citep{1982Icar...49..298C,2012ApJ...756...90V}. Once available from dedicated evolutionary simulations, we will use evolution tracks including irradiation in future iterations of the global model.

\subsection{Mass loss}\label{sec:massloss}

When a clump moves closer to the star, its Hill radius $R_\mathrm{H}$ decreases.
At some point, the clump's physical radius given by the tracks can become larger than its Hill radius.
We assumed that in this case the clump  loses the mass outside of its Hill sphere (Roche lobe overflow).
In our model, $R_\mathrm{c}$ is compared with $R_\mathrm{H}$ at each time step.
If $R_\mathrm{c} > R_\mathrm{H}$, the clump's mass is reduced by a small amount and $R_\mathrm{c}$ is redetermined from the evolution tables.
The new radius is again compared with $R_\mathrm{H}$ for the reduced mass.
This process is repeated until $R_\mathrm{c} \le R_\mathrm{H}$.
In some cases, no stable configuration is found, which implies that the clump was entirely disrupted (tidal destruction).
Any mass lost in this process is returned to the disc at the clump's location via a source term ($S_\mathrm{loss}$ in Eq.~\ref{eq:evo}). The same happens if a clump is destroyed due to thermal disruption (Sect.~\ref{sec:irrad}).

\section{Clump-disc and clump-clump interactions}
\label{sec:interact}
\subsection{Gas disc-driven orbital migration }\label{sec:migration}

A companion embedded in a circumstellar disc interacts gravitationally with the disc gas. The  exchange of angular momentum leads to orbital migration \citep{1980ApJ...241..425G}.
Classically, gas disc migration has been divided into two main regimes. In the type~I regime, the surface density remains largely unperturbed by the presence of the companion. In a non-self-gravitating disc, this is the case for planets up to a few to a few tens of Earth masses ($\mearth$).
The exact value depends on the stellar mass and disc properties.
Type~I migration is complex and is strongly affected by the disc's thermodynamics. Some of the contribution to the total torque comes from Lindblad resonances and is typically negative (inwards) \citep{goldreichward1973}. Additional contributions come from the co-rotation region and these can be both negative and positive, such that the net migration can be both inwards and outwards \citep{2006A&A...459L..17P}.
An important aspect in determining whether outward migration is possible is the saturation of the co-rotation torque at higher masses, since the saturation works against outward migration \citep{Paardekooper2011b}.
More massive companions start to perturb the surface density significantly until they clear a gap in the vicinity of their orbit. They are then said to migrate in type~II.
The simple picture consisting of type~I and type~II migration is helpful when studying planets formed in CA. Such planets are assembled from initial masses of $\ll \SI{1}{\mearth}$ that need to accrete mass for many orbits to eventually open a gap.
Prescriptions for the type~I migration (e.g. \citealt{2002ApJ...565.1257T,Paardekooper2010,Paardekooper2011b}) can then be applied until some criterion for gap opening is fulfilled \citep{2006Icar..181..587C,2018ApJ...861..140K}.
Above gap-opening masses, type~II prescriptions can then be applied \citep{2014A&A...567A.121D,2018ApJ...861..140K}.

In DI, the situation is different. Fragments are born with high masses $\sim\SI{}{\mj}$ and they may migrate too fast to carve a deep gap despite their high masses.
Furthermore, migration is expected to be to some degree stochastic \citep{2011MNRAS.416.1971B,Kubli2025}.
The conventional formulae for type~I migration assuming a steady state cannot be applied, and the  conditions for gap opening are still uncertain \citep{2015ApJ...802...56M,2018ApJ...854..112M,2020MNRAS.496.1598R}.
Furthermore, while the transfer of angular momentum from a $\sim \SI{}{\mearth}$ planet on the disc gas may be negligible, this is  not the case for a clump of several $\mj$.

\subsubsection{Torque densities}\label{sec:torquedens}
In order to overcome these limitations, we applied the migration model from \citetalias{2022A&A...664A.138S}. 
The migration model uses torque densities to account for the two-way exchange of angular momentum between the disc and the companions.
The torque density distribution per unit disc mass, $\frac{d\mathcal{T}}{dm}(r)_i$, for companion $i$ is defined such that 
\begin{equation}\label{eq:torquedens}
    \mathcal{T}_i = 2 \pi \int_0^\infty\!\left(\frac{\mathrm{d}\mathcal{T}}{\mathrm{d} m}\right)_i(r)\Sigma(r) r\,\mathrm{d} r,
\end{equation}
where $\mathcal{T}_i$ is the total torque on companion $i$: the sum of the contributions from all radial locations in the disc.
In the evolution equation (Eq.~\ref{eq:evo}) we thus have for $n$ companions:
\begin{equation}
    \Lambda = \sum_{i=1}^{n} - \left(\frac{\mathrm{d}\mathcal{T}}{\mathrm{d} m}\right)_i(r).
\end{equation}
In principle, this formalism fully integrates the exchange of angular momentum between the disc and the companions in our model.
Hereafter, we drop the subscript $i$ and consider the case of an individual planet for readability.

We note, however, that two important questions remain unanswered regarding
(1)  the shape of $\frac{d\mathcal{T}}{dm}(r)$ and 
(2) what happens with planets on eccentric orbits.

For the first question, possible choices are the classical impulse approximation \citep{1979MNRAS.188..191L,1979MNRAS.186..799L,1986ApJ...309..846L} or an improved formalism \citep{2002MNRAS.334..248A}. These approaches can work reasonably well in a pure type~II regime where a gap has already been fully opened. However, they do not work well in type~I or in the transition between type~I and type~II because the approach explicitly excludes co-rotation torques. As discussed above, co-rotation torques are important there.
Fortunately, torque densities are well suited for the type~I regime as well -- provided they include co-rotation torques.
A model that accomplishes this is presented in \citet{2010ApJ...724..730D}. The authors performed 3D nested grid hydrodynamic simulations of locally isothermal discs. They provided a functional form of the torque density distribution parametrised by the disc's surface density and temperature gradients,
\begin{equation}\label{eq:tdens}
    \frac{\mathrm{d} \mathcal{T}}{\mathrm{d} m}(r) = \mathcal{F} \left(x, \beta, \zeta\right) \Omega(r_\mathrm{c})^2 r_\mathrm{c}^2 \left(\frac{M_\mathrm{c}}{\mstar}\right)^2 \left(\frac{r_\mathrm{c}}{H(r_\mathrm{c})}\right)^4,
\end{equation}
where 
\begin{equation}\label{eq:f}
    \mathcal{F}(x,\beta,\zeta) = \left[ p_1 e^{-(x+p_2)^2/p_3^2} + p_4 e^{-(x-p_5)^2/p_6^2} \right] \tanh{(p_7-p_8 x)}
\end{equation}
is a dimensionless function describing the shape of the torque density.
In Eq.~\ref{eq:f}, $\beta = - \mathrm{d} \ln \Sigma / \mathrm{d} \ln r$ and $\zeta = - \mathrm{d} \ln T / \mathrm{d} \ln r$, $x = \frac{r-r_\mathrm{c}}{\mathrm{max}(H,R_\mathrm{H})}$ is a scaling factor. The parameters $p_1$ to $p_8$ depend on $\beta$ and $\zeta$ and set the amplitude and width of the torque density. They  are determined by \citet{2010ApJ...724..730D} via fits to their simulations.
This form of the torque density is adequate to the type~I regime, \citet{2010ApJ...724..730D} study \SI{1}{\mearth}-planets in most of their simulations. However, they also explored what happens in the case of gap-opening planets.
They showed that when the planetary mass increases, the amplitude (scaled) torque density decreases, although only by a factor of order unity.

\subsubsection{High mass torque}\label{sec:hmt}
In addition to the decreasing amplitude, as the planetary  mass approaches \SI{1}{\mj}, an inversion of the torque density starts to appear near the position of the planet (their Fig.~15).
The exact shape of this inversion, which depends strongly on the planet's mass is, however, not very important. What is important is that the torque is small near the planet.
In \citetalias{2022A&A...664A.138S} we studied this effect by applying a torque model named the 'high-mass torque model' resulting from an interpolation of Fig.~15 in \citet{2010ApJ...724..730D} to accreting, migrating planets.
We argued that the inversion of the torque density prevents premature gap opening and compared the migration tracks of these 1D simulations with the 3D nested grid hydrodynamic simulations of \citetalias{2008ApJ...685..560D}, finding reasonable agreement for a range of parameters.
We also showed that without the modifications related to the planet's mass, a gap opens too fast and migration slow down too quickly. 
While the torque model performs well in comparison with locally isothermal 3D simulations, it cannot reproduce features such as the outward migration that depend on the (non\nobreakdash-)saturation of the co-rotation torque.
We propose a simple method to overcome this limitation in Appendix~F of \citetalias{2022A&A...664A.138S}.
For the present study, these effects should not be important, since we are dealing with massive objects for which the co-rotation torque saturates.
Therefore, we applied the high mass torque from \citetalias{2022A&A...664A.138S} in Eq.~\ref{eq:torquedens}.
Since we also applied this shape of the torque density to higher-mass companions, we prevented the inversion close to the companion by setting $\frac{\mathrm{d} \mathcal{T}}{\mathrm{d} m} =0$ near the companion \citep{2017MNRAS.469.3813H}.
We note that despite our treatment of the torque densities at high masses, their application to 1D discs leads to gap shapes that are different from what is seen in hydrodynamic simulations.
This is seen in simulations running for hundreds of orbits and the gaps in 1D tend to become too deep and too narrow.
We discuss this issue and its possible consequences in Sect.~\ref{sect:disc_torque} and will address this topic in future research. 

\subsubsection{Eccentric and inclined orbits}
The second question raised in Sect.~\ref{sec:torquedens} about companions on eccentric orbits is particularly  important for multiple systems.
Even if all companions (like newly formed clumps) were initially on (approximately) flat, circular orbits, the mutual gravitational interaction of the companions tends to increase their eccentricity and inclinations.
Furthermore, in the context of DI, we expect the clumps to be on initially eccentric orbits, since they are born near the corotation of spiral arms.

Strictly speaking, the torque density formulation discussed in Sect.~\ref{sec:torquedens} is only applicable for nearly circular, flat orbits.
The same is true for the type~I migration formulae mentioned in Sect.~\ref{sec:migration}. These must be  modified \citep{2008A&A...482..677C,Bitsch2010,2014MNRAS.437...96F,Coleman2014}.
The treatment in the context of torque densities is not trivial.

To our knowledge, no published torque densities for eccentric, inclined orbits exist (but see \citealt{Fairbairn2025}).
However, simply applying, for example, the torque density corresponding to a companion's semi-major axis to the disc when the companion is on a strongly eccentric orbit  violates the conservation of angular momentum.
Numerical experiments show that doing so tends to add angular momentum to the system, which can lead to unphysical fragmentation events and other complications.
We stress that such difficulties arise only if the companion's torque is applied on the disc through Eq.~\ref{eq:evo}.
An additional complication arises in relation to the damping of eccentricities and inclinations.
This damping is expected as long as the companions are embedded in a gas disc. For massive companions in type~II migration, hydrodynamic simulations show that damping occurs on timescales shorter than the migration timescale \citep{2004A&A...414..735K,2019SAAS...45..151K}.
In this situation, the so-called $K$-damping is often applied \citep{2002ApJ...567..596L}, where the damping timescale is set to $1/K$ times the migration timescale.
$K$ is typically of the order of ten \citet{2021A&A...656A..69E}.
The damping is then applied to the companion as an additional acceleration in the N-body integrator.
In the context of our model, this becomes problematic, since it means applying a torque or force to a companion without applying the opposite equal to the disc.
Again, this leads to unphysical effects in the disc evolution.

For this reason, we developed a more robust way to model migration and damping for massive companions.
It is based on the formulation of planet-disc interaction in the framework of dynamical friction presented in \citet{2020MNRAS.494.5666I}.
The authors studied planet-disc interactions in the subsonic and supersonic regimes and compared and discussed in detail various approaches to calculate damping found in the literature.
Based on this discussion and on the results of a hydrodynamic simulation, they provided a simple prescription for the damping timescales of semi-major axis, eccentricity and inclination for low-mass planets.
In order for this approach to be applicable in our model, we need to demonstrate that, in the limit or circular orbits, we retrieve the same torque as the one obtained from the high mass torque (Sect.~\ref{sec:hmt}).
Furthermore, it also needs to be applicable for massive companions.

\citet{2020MNRAS.494.5666I} assumed discs with smooth surface densities in their Sect.~4 to arrive at equations of motion that can be expressed in terms of the migration timescale (their Eq.~46). In order to apply the dynamical friction formalism to gap-opening companions, we need to relax this assumption.
Instead, we start from their Eqs.~29~and~32. Their Eq.~29 gives the (specific) force from dynamic friction in the subsonic limit \citep{2004ApJ...602..388T}:
\begin{equation}
    \vec{F}_\mathrm{DF,sub} = -0.780 \frac{\Delta \vec{v}}{t_\mathrm{wave}},
\end{equation}
with the characteristic (inverse) timescale of
\begin{equation}
    t_\mathrm{wave}^{-1} = q \frac{\Sigma_\mathrm{c} r_\mathrm{c}^2}{\mstar} h^{-4} \Omega_\mathrm{c},
\end{equation}
we have the mass ratio of the companion to the star,
\begin{equation}
    q = \frac{M_\mathrm{c}}{\mstar},
\end{equation}
the disc aspect ratio,
\begin{equation}
    h = \frac{H(r_\mathrm{c})}{r_\mathrm{c}},
\end{equation}
the torque
\begin{equation}
    \Gamma_\mathrm{o} = \left(\frac{q}{h}\right)^2 \Sigma_\mathrm{c} r_\mathrm{c}^4 \Omega_\mathrm{c}^2, 
\end{equation}
and the velocity of the companion relative to the gas,
\begin{equation}
    \Delta \vec{v} = \vec{v}_\mathrm{c} - \vec{v}_\mathrm{g},
\end{equation}
where: 
\begin{equation}
    \vec{v}_\mathrm{c} = \left(\begin{array}{c} v_\mathrm{c,r} \\ v_\mathrm{c,\theta} \\ v_\mathrm{c,z}, \end{array}\right), 
\end{equation}
and, assuming an unpertured disc velocity field,
\begin{equation}
    \vec{v}_\mathrm{g} = \left(\begin{array}{c} v_\mathrm{g,r}(r_\mathrm{c}) \\ r_\mathrm{c} \Omega_\mathrm{g} (r_\mathrm{c}) \\ 0 \end{array}\right). 
\end{equation}
These are the velocities of the companion and the gas (at the companion's separation), respectively, in cylindrical coordinates.
For $\Delta v \equiv |\Delta \vec{v}|$ we get
\begin{equation}
    \Delta v = \sqrt{(v_\mathrm{c,r}-v_\mathrm{g,r})^2+(v_\mathrm{c,\theta} - r \Omega_\mathrm{g})^2+v_\mathrm{c,z}^2}.
\end{equation}
The angular frequency of the gas, $\Omega_\mathrm{g}$ is given by 
\begin{equation}
    \Omega_\mathrm{g} = \Omega (1-\eta),
\end{equation}
with $\Omega$ from Eq.~\ref{eq:angfreq}.
The factor 
\begin{equation}\label{eq:eta}
    \eta \equiv - \frac{h^2}{2} \frac{\mathrm{d} \ln p}{\mathrm{d} \ln r}
\end{equation}
accounts for the deviation of the gas velocity from the purely gravitational value due to the pressure gradient. For smooth discs this effect is very small, but it is important here because otherwise the $\theta$-component of $\Delta \vec{v}$ vanishes for a companion on a circular orbit.
For the supersonic limit, \citet{2020MNRAS.494.5666I} apply the expression from \citet{2011ApJ...737...37M} (their Eq.~32, simplified):
\begin{equation}
    \vec{F}_\mathrm{DF,sup} \simeq - 2 \pi \frac{\Delta \vec{v}}{t_\mathrm{wave}} \left( \frac{\Delta v}{c_\mathrm{s}} \right)^3.
\end{equation}
Summation of timescales then leads to (as in their Eq.~35):
\begin{equation}\label{eq:df}
    \vec{F}_\mathrm{DF} = \frac{\Delta \vec{v}}{\Delta v} \left[ \frac{1}{-0.78 \Delta v t_\mathrm{wave}^{-1}} + \frac{1}{-2\pi \Delta v t_\mathrm{wave}^{-1}}\left( \frac{\Delta v}{c_\mathrm{s}} \right)^3 \right]^{-1}.
\end{equation}
As discussed in \citet{2020MNRAS.494.5666I}, the dynamical friction formalism can recover the migration rate in the subsonic case, but only to an order of magnitude. This is because in the subsonic regime, the contribution by density waves is important.
A way to include this contribution without assuming low masses is to modify Eq.~\ref{eq:df}.
To do this, a relationship between the total torque calculated through Eq.~\ref{eq:torquedens} and $t_\mathrm{wave}^{-1}$ is needed.
We write the total torque on a companion as:
\begin{equation}
    \mathcal{T} = \widetilde{K}~\Gamma_\mathrm{o},
\end{equation}
where $\widetilde{K}$ is a dimensionless prefactor.
With the migration timescale, $\tau$:
\begin{equation}
    \tau^{-1} = \frac{2 \mathcal{T}}{L_\mathrm{c}} = \frac{2 \widetilde{K} \Gamma_\mathrm{o}}{M_\mathrm{c} r_\mathrm{c}^2 \Omega_\mathrm{c}},
\end{equation}
where 
\begin{equation}\label{eq:lp}
    L_\mathrm{c} = M_\mathrm{c} r_\mathrm{c}^2 \Omega_\mathrm{c}, 
\end{equation}
is the orbital angular momentum of the companion, which can be expressed as
\begin{equation}
    t_\mathrm{wave}^{-1} = \frac{1}{2 \widetilde{K} h^2} \tau^{-1} = \frac{\Gamma_\mathrm{o}}{L_\mathrm{c} h^2}.
\end{equation}
If we assume $\Sigma(r) \sim \Sigma_\mathrm{c}$ in Eq.~\ref{eq:torquedens} and substitue from Eq.~\ref{eq:tdens}, then formally we can write: 
\begin{equation}
    \widetilde{K} = 2 \pi H^{-2} \int_{0}^{\infty}\!\mathcal{F} \left(\frac{r-a}{H},\beta,\zeta\right) r\,d r.
\end{equation}
In summary, we have
\begin{equation}\label{eq:sum}
\begin{aligned}
    \mathcal{T} &= \widetilde{K} \Gamma_\mathrm{o} = \widetilde{K} L_\mathrm{c} h^2 t_\mathrm{wave}^{-1}\\
    &= 2 \pi H^{-2} L_\mathrm{c} h^2 t_\mathrm{wave}^{-1} \int_{0}^{\infty}\!\mathcal{F} \left(\frac{r-a}{H},\beta,\zeta\right) r\,d r\end{aligned} 
\end{equation}
Now Eq.~\ref{eq:df} can be modified. 
We replace the factor $-0.78$ by $\widetilde{K} h^2 \eta^{-1}$.
In order to see that this choice is reasonable, we consider the $\theta$-component for a circular orbit. Then we have $\Delta v / c_\mathrm{s} << 1$ and $\Delta v = v_\mathrm{c,\theta} - r_\mathrm{c} \Omega_\mathrm{g} = r_\mathrm{c} \eta \Omega$, and thus:
\begin{equation}
\begin{aligned}
    F_\mathrm{DF,\theta} &= \left[ \frac{\eta}{\widetilde{K} h^2 r_\mathrm{c} \eta \Omega t_\mathrm{wave}^{-1}} \right]^{-1}\\
    &= \frac{\widetilde{K} h^2 r_\mathrm{c}^2 M_\mathrm{c} \Omega_\mathrm{c} t_\mathrm{wave}^{-1}}{M_\mathrm{c} r_\mathrm{c}}\\
    &= \frac{\widetilde{K} L_\mathrm{c} h^2 t_\mathrm{wave}^{-1}}{M_\mathrm{c} r_\mathrm{c}}\\
    &= \frac{\mathcal{T}}{M_\mathrm{c} r_\mathrm{c}}.
\end{aligned}
\end{equation}
In the first step, we use $\Omega =\Omega_\mathrm{c}$ for a circular orbit. In the second step, we substitute the expression from Eq.~\ref{eq:lp} and in the third, we drew from Eq.~\ref{eq:sum}.
Thus, we retrieve the acceleration of the companion when subject to the torque from Eq.~\ref{eq:torquedens}.
In the general case, we therefore apply the following acceleration (see Sect.~\ref{sec:nbody} for the sign of $\mathcal{T}$):
\begin{equation}\label{eq:ftddf}
    \vec{F}_\mathrm{TD,DF} = \frac{\Delta \vec{v}}{\Delta v} \left[ \frac{M_\mathrm{c} r_\mathrm{c}}{\mathcal{T}} - \frac{1}{2\pi \Delta v t_\mathrm{wave}^{-1}}\left( \frac{\Delta v}{c_\mathrm{s}} \right)^3 \right]^{-1}.
\end{equation}
It includes the contribution to migration by density waves, is not limited to smooth discs and reproduces the supersonic limit.

Equation~\ref{eq:ftddf} also illustrates what we have discussed above: in case of circular orbits, the torques on companion and disc are equal but opposite. Conversely, on eccentric orbits, $v_\mathrm{c,\theta}$ is typically different from $r \Omega_\mathrm{g}$. Therefore, it would not be correct to apply the torque from EQ.~\ref{eq:torquedens} to the disc.
Instead, when applying the torque to the disc, it is scaled by the sine of the angle between the position vector $\vec{r}_\mathrm{c}$ and the velocity vector, $\vec{v}_\mathrm{g}$,
\begin{equation}
    \mathcal{T}_\mathrm{corr} = \mathcal{T} \frac{\vec{r}_\mathrm{c} \times \vec{v}_\mathrm{g}}{|\vec{r}_\mathrm{c}| |\vec{v}_\mathrm{g}|}.
\end{equation}
This ensures that the torque on the disc is reduced for eccentric orbits and gives the correct limit for radial motion (zero torque on the disc).

We note that our treatment of synchronising the torques between the companion and the disc fully accounts for accelerations only in $\theta$ direction.
Applying torques in $r$- and $z$- direction would make the disc locally eccentric/inclined. This is not possible in our model which corresponds to a rotationally symmetric disc centred in the $z=0$ plane.

\subsection{Gravitational interaction}\label{sec:nbody}

In addition to the tidal interaction with the gas disc, companions also interact with each other gravitationally.
We employ the \texttt{mercury} N-body code \citep{Chambers1999} to model such interactions. It is a symplectic integrator that applies a hybrid method in order to correctly handle close encounters.
\texttt{mercury} is called in each main time step.
Since the companion positions are updated in the N-body code, the result of the companions' interaction with the gas disc must also be included there.
This is done by means of additional (specific) forces.
The form of these forces related to gas disc migration as well as $i$- and $e$-damping is discussed in Sect.~\ref{sec:migration}.
Here we give the final expressions that are passed to the N-body code.
There is a  subtlety to the application of these forces.
It is done in cylindric, heliocentric coordinates. This means that the companions' positions and velocities first need to be transformed from democratic to heliocentric coordinates, and then decomposed into tangential and radial components.
After the application, the transformation needs to be inverted.
The force in $r$-direction is
\begin{equation}\label{eq:f_r}
    F_\mathrm{r} = \frac{v_\mathrm{c,r} + \frac{\dot{m}}{2 \pi r \Sigma_\mathrm{c}}}{\Delta v} \left[- \frac{M_\mathrm{c} r_\mathrm{c}}{|\mathcal{T}|} - \frac{1}{2\pi \Delta v t_\mathrm{wave}^{-1}}\left( \frac{\Delta v}{c_\mathrm{s}} \right)^3 \right]^{-1} - \frac{d}{dr} V_\mathrm{d}(r).
\end{equation}
The first summand in Eq.~\ref{eq:f_r} is almost as in the $r$-component of Eq.~\ref{eq:ftddf}, (we substituted $-\frac{\dot{m}}{2 \pi r \Sigma_\mathrm{c}}$ for $v_\mathrm{g,r}$, with $\dot{m} \equiv \dot{m}(r)$ the gas accretion rate in the disc).\footnote{In our convention, $\dot{m}$ is positive when gas is moving towards the star (accretion). In this case, and if $v_\mathrm{r} > 0$, the relative velocity between companion and gas is increased.}
The difference is the sign of the first summand in the square brackets. This term is always negative since it corresponds to the damping, which needs to be in the direction opposite to the companion's relative motion.
The second summand in Eq.~\ref{eq:f_r} is the additional acceleration resulting from the disc's gravitational potential.
In $\theta$-direction we have the force:
\begin{equation}\label{eq:f_theta}
    F_\mathrm{\theta} = \frac{v_\mathrm{c,\theta} -r_\mathrm{c} \Omega_\mathrm{g}(r_\mathrm{c})}{\Delta v} \left[ \frac{M_\mathrm{c} r_\mathrm{c}}{\mathcal{T}} - \frac{1}{2\pi \Delta v t_\mathrm{wave}^{-1}}\left( \frac{\Delta v}{c_\mathrm{s}} \right)^3 \right]^{-1},
\end{equation}
exactly as in the $\theta$-component of Eq.~\ref{eq:ftddf}.
Migration is typically directed inward ($\mathcal{T}<0$), although it can also become positive.
For the $z$-direction, we  obtained
\begin{equation}\label{eq:f_z}
    F_\mathrm{z} = \frac{v_\mathrm{c,z}}{\Delta v} \left[- \frac{M_\mathrm{c} r_\mathrm{c}}{|\mathcal{T}|} - \frac{1}{2\pi \Delta v t_\mathrm{wave}^{-1}}\left( \frac{\Delta v}{c_\mathrm{s}} \right)^3 \right]^{-1} - \mathrm{sign}(z_\mathrm{c}) 2 \pi G \Sigma_\mathrm{enc}, 
\end{equation}
with the same sign in the $\mathcal{T}$-term as in Eq.~\ref{eq:f_r}.
$z_\mathrm{c}$ is the companion's $z$-coordinate.
In Eq.~\ref{eq:f_z}, the last summand is the acceleration an inclined companion experiences due to the disc's gravity.
We assumed that this acceleration results from an infinitely extended, thin disc with the surface density $\Sigma_\mathrm{enc}$ enclosed by the companion.
We note that this additional term does not damp the inclination of the companion, since it is directed against $v_\mathrm{c,z}$ only if $z_\mathrm{c}$ and $v_\mathrm{c,z}$ have the same sign.  

\subsection{Collisions}
When the companions interact gravitationally they can experience close encounters, leading to scatterings, or collisions.
In this study, we treated all collisions as perfect mergers: when two companions come closer than the sum of their radii, a perfectly inelastic collision is assumed to occur.
This may appear as an extreme assumption since it neglects the possibility of eruptive or hit-and-run collisions that would change the orbits of the bodies in question without completely destroying one of them. 
This is especially true for clumps due to their large extension. Indeed, it was shown that collisions between clumps can lead to various outcomes depending on the assumed conditions (see \citealt{Matzkevich2024} for details).
However, we found that most clump-clump collision typically occur below the mutual escape speed.
In this regime (see also \citet{Wimarsson2025} for a similar collisional regime in a different context), perfect merging may be a reasonable assumption (e.g. \citealp{2012ApJ...745...79L,2021MNRAS.502.1647C}).
We study collisions further in \citetalias{Schib2025b}.

Overall, the global disc instability formation model presented here contains - in the low dimensional approximation - a significant number of physical processes that are coupled in a self-consistent way. At the same, it is still a strong simplification of the actual companion formation and evolution process. We discuss important limitations in Sect. \ref{sect:Discussion}.

\section{Formation of individual systems}\label{sec:sys}
Thanks to its low-dimensional character, it is possible to employ the global model for population syntheses. However, before studying the result of the model in a statistical sense in \citetalias{Schib2025b}, it is important to discuss the evolution of several individual systems, studying outcomes of increasing complexity. 
Thus, we began by looking at the simple case of a system that does not fragment.
Then we studied a system with a single fragmentation event that only produces one clump.
Next, we looked at a system with two fragmentation events.
Finally, we investigated a more complex system evolution with multiple fragmentation events and several interacting clumps.

\subsection{A non-fragmenting system}\label{sec:nonfrag}
\begin{figure*}
    \begin{subfigure}[pt]{0.49\textwidth}
    \includegraphics[width=\linewidth]{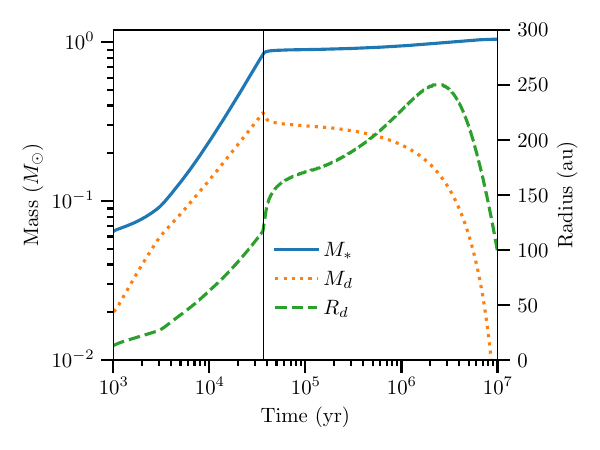}
    \end{subfigure}
    \begin{subfigure}[pt]{0.49\textwidth}
    \includegraphics[width=\linewidth]{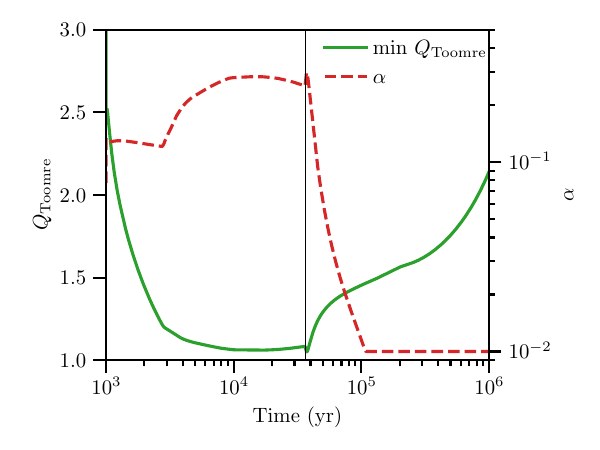}
    \end{subfigure}
    \caption{Evolution of a system that becomes self-gravitating but that does not fragment. 
    Left panel: Stellar mass, disc mass and disc size as a function of time.
    Right panel: Minimum $Q_\mathrm{Toomre}$ and $\alpha$ as a function of time. The black vertical line shows the end of infall. It should be noted that there two distinct y-axes in both panels to which the different quantities belong to.}
    \label{fig:nonfrag}
  \end{figure*}

Figure~\ref{fig:nonfrag} shows the time evolution of a system where the disc becomes self-gravitating (i.e. where $Q_\mathrm{Toomre}$ approaches unity and spiral waves would form)  but where no fragmentation occurs. As we shall see in \citetalias{Schib2025b}, such an outcome is the most probable one: while almost all discs are found to have early on a self-gravitating phase, only 10-20\% fragment. Thus, only the disc part of the model is used. 
This system is initialised at \SI{1}{kyr} with a stellar seed of $\approx \SI{6.5e-2}{\msun}$ and a disc of ($\approx \SI{2e-2}{\msun}$).
It experiences infall at $\approx \SI{3e-5}{\msun yr^{-1}}$ until $t \simeq \SI{36}{kyr}$.
The left panel of Fig.~\ref{fig:nonfrag} displays the stellar and disc masses as a function of time.

The disc mass is initially less than a third of the stellar mass.
Due to the rapid infall, the disc mass increases quickly to two thirds the stellar mass.
This leads to a very fast transport of angular momentum via the global instability of the disc (Sect.~\ref{sec:global}).
The associated $\alpha$ parameter is shown in the right panel of the figure.
It reaches almost \num{0.3} in the early phase.
The $Q_\mathrm{Toomre}$ parameter, also shown in the right panel, drops to low values, but remains just above unity.
After the infall phase terminates (indicated with a thin vertical line), $\alpha$ drops quickly, although it remains elevated above background $\alpha_\mathrm{BG}$ for more than \SI{100}{kyr}.
This coincides well with the overall minimum value of $Q_\mathrm{Toomre}$, which is lower than about \num{1.5} in the same period.
We expect such a system to exhibit spiral arms for about \SI{100}{kyr}. Here, we see the typical auto-regulation of self-gravitating discs: as $Q_\mathrm{Toomre}$ approaches unity, spiral waves develop which efficiently transport away mass and angular momentum. In our model, this is expressed by the increase of $\alpha$ reaching a high value of almost 0.3. This increases in turn $Q_\mathrm{Toomre}$ via a higher disc temperature and reduced surface density. In this system, fragmentation is prevented this way.  It is also a typical behaviour that the overall lowest $Q_\mathrm{Toomre}$ is reached directly after the end of infall, because at this moment, the heating (and thus stabilisation) of the disc by shock heating of the disc's surface by infalling gas vanishes. 

\subsection{A simple fragmenting system}\label{sec:sfrag}
\begin{figure*}[htb!]
    \begin{subfigure}[pt]{0.49\textwidth}
        \includegraphics[width=\linewidth]{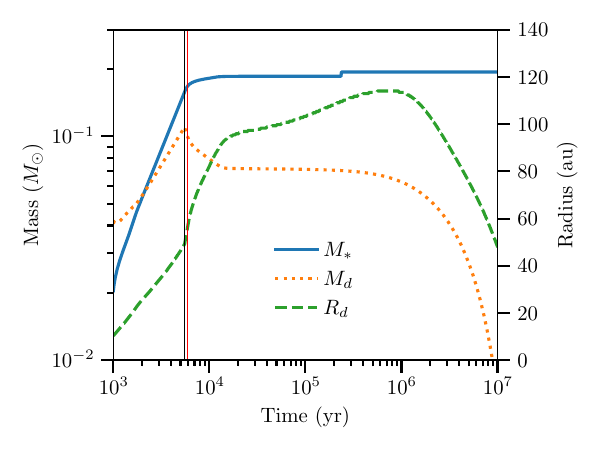}
    \end{subfigure}
    \begin{subfigure}[pt]{0.49\textwidth}
        \includegraphics[width=\linewidth]{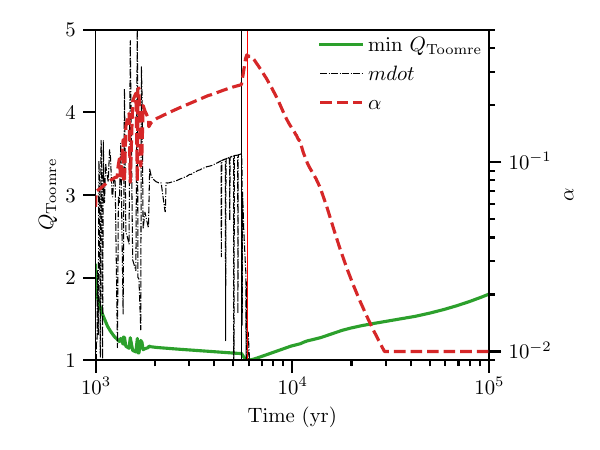}
    \end{subfigure}
    \begin{subfigure}[pt]{0.49\textwidth}
        \includegraphics[width=\linewidth]{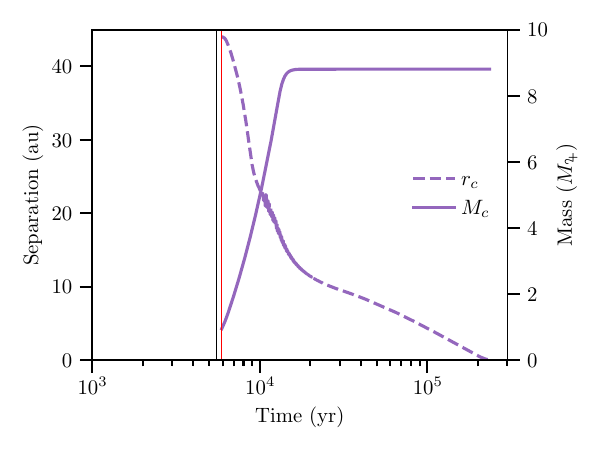}
    \end{subfigure}
    \begin{subfigure}[pt]{0.49\textwidth}
        \includegraphics[width=\linewidth]{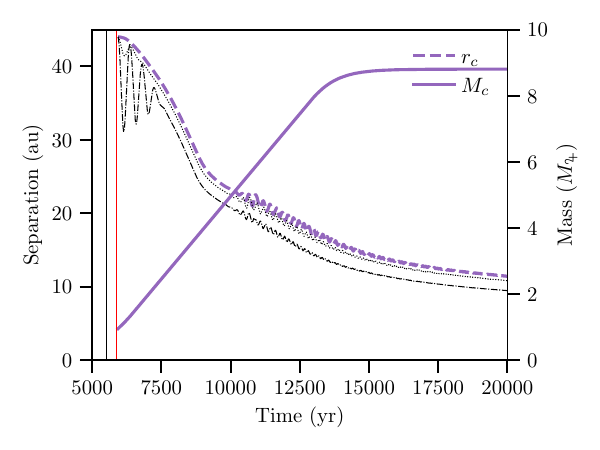}
    \end{subfigure}
    \begin{subfigure}[pt]{0.49\textwidth}
        \includegraphics[width=\linewidth]{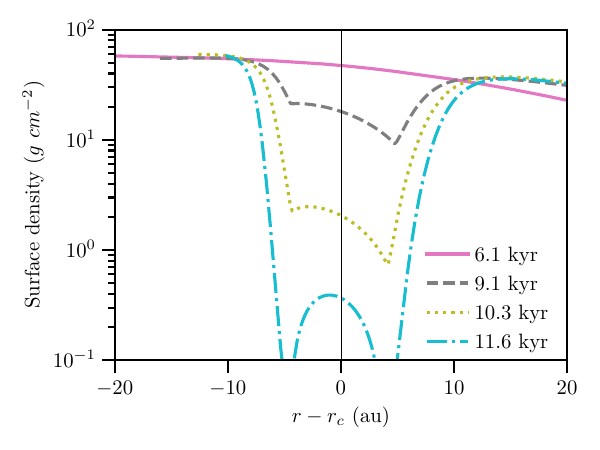}
    \end{subfigure}
    \begin{subfigure}[pt]{0.49\textwidth}
        \includegraphics[width=\linewidth]{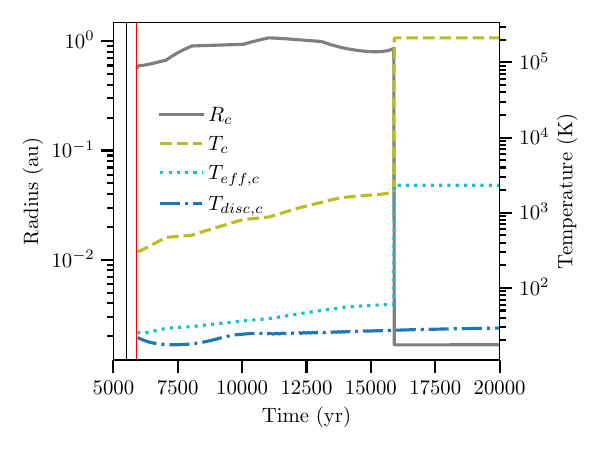}
    \end{subfigure}
    \caption{Evolution of a system forming a single (transient) fragment which is eventually accreted into the host star because of orbital migration.
    Top left: Stellar mass, disc mass and disc radius.
    Top right: Minimum $Q_\mathrm{Toomre}$ and $\alpha$. The stellar accretion rate is shown in black (see text Sect.~\ref{sec:sfrag}).
    Middle left: Time evolution of fragment orbital separation $r_{\rm c}$ and mass $M_{\rm c}$.
    Middle right: Same with a zoom-in on the early phase, two variant calculations are shown additionally in black (see text Sect~\ref{sec:sfrag}).
    Bottom left: Surface density, centred on the fragment (and thus at different absolute orbital distances), for a selection of times.
    Bottom right: Time evolution of the fragment radius, central and  effective temperature as well as the disc temperature at the fragment's location.}
    \label{fig:sfrag}
\end{figure*}
The next system we investigate has a lower total mass than the one discussed above, although the behaviour of stellar mass and disc mass is qualitatively similar (top left panel of Fig.~\ref{fig:sfrag}).
The initial values are \SI{0.02}{\msun} and \SI{0.04}{\msun} for stellar mass and disc mass, respectively, with infall at \SI{4.9e-5}{\msun yr^{-1}} until \SI{5.5}{kyr}.
In this case (with a very high disc-to-star mass ratio) the unstable initial state leads to accretion bursts. This is seen as an oscillatory feature in $Q_\mathrm{Toomre}$ and $\alpha$.
We checked that these oscillations are not numerical in nature by rerunning the simulation at a much lower time step.
The thin black dash-dotted line in the top right panel of the figure, shows the accretion rate of disc material on the star $mdot$ (in units of \SI{e-4}{\msun yr^{-1}} vs the left y-axis).
The accretion rate also shows strong oscillations between \num{1} and \SI{2}{kyr}.
The final stellar mass is $\approx \SI{0.2}{\msun}$.

In Fig.~\ref{fig:sfrag}, the end of the infall phase is again marked with a black vertical line.
In this case, given the combination of quantities setting $Q_\mathrm{Toomre}$, a value slightly less than unity is reached and a fragment is formed with an initial mass (set according to Eq.~\ref{eq:mfraginit} based on the local disc conditions) of \SI{0.9}{\mj} at \SI{44}{au}.
The middle left panel shows the time evolution of mass (solid line) and separation (dashed line) of the fragment.
The red vertical line at $\approx \SI{6}{kyr}$ marks the fragmentation event.
Initially, the fragment migrates inwards rapidly, while accreting gas from the disc.
At approximately \SI{10}{kyr} (or after about six initial orbits), the migration is slowed down.
At this time, the forming companion has already accreted almost \SI{4}{\mj} of gas.
The middle right panel shows a zoom-in on the same data with a linear x-axis, revealing more details of what is happening in the early phase.
The accretion rate is at its maximum of \SI{e3}{\mj.yr^{-1}} during the early phase due to the massive disc and the fast transport.
The rapid inward migration is slowed down after $\approx \SI{10}{kyr}$ with the onset of the formation of a gap.

The fragment in Fig.~\ref{fig:sfrag} starts with a moderate initial angular momentum deficit (corresponding to an eccentricity of $\approx \num{0.08}$, see Sect.~\ref{sec:auto}) and does not exhibit strong oscillations in its separation at \SI{6}{kyr} (middle right panel). 
To understand the role of different initial orbital eccentricities, we run two variant simulations which only differs by the  initial eccentricity. They are shown only in the middle right panel. In the first variant simulation, we studied how the fragment would behave with a higher initial `eccentricity' (see Sect.~\ref{sec:auto}) of \num{0.25}, shown by the thin dotted line in the middle right panel.
The fragment starts from apocentre and initially moves inwards quickly, but is essentially circularised through damping in a single orbit. The difference in separation after \SI{5}{kyr} is less than \SI{1}{au}.
In the second variant simulation (dashed black line), a fragment with double that eccentricity (corresponding to 0.5, higher than the maximum used for the population synthesis) is inserted.
In this case, circularisation happens in about three orbits and the fragment has moved $\approx$~\SI{2}{au} further inwards (at \SI{25}{kyr}) compared the one with lowest angular momentum deficit.
This indicates that the initial eccentricity does not have a major impact on companions formed by DI, at least in the single fragment case.
We discuss this further in \citetalias{Schib2025b}.

During the first $\SI{3}{kyr}$ after fragmentation, the disc  receives only little perturbation from the fragment, which migrates quickly.
Later, with increasing fragment mass, the disc material is pushed away more vigorously.
After $\approx$~\SIrange{4}{6}{kyr}, the combination of tidal interaction and gas accretion has led to a deep gap.
This process is shown in the bottom left panel of Fig.~\ref{fig:sfrag}.
It shows the surface density centred on $r-r_\mathrm{c}$ for several snapshots in the early migration.
With the formation of the gap, the accretion rate starts to decrease due to the depletion of the disc near the companion.

The bottom-right panel in Fig.~\ref{fig:sfrag} finally displays the fragment's internal properties, namely its radius, central temperature, $T_\mathrm{c}$, and  the effective temperature, $T_\mathrm{eff,c}$.
The radius of the clump increases from an initial \SI{0.6}{au} to \SI{1}{au} under the combined influence of accretion and contraction.
The central temperature riises from $\approx \SI{300}{K}$ to $\approx \SI{2000}{K}$. At this point, the clump undergoes a dynamical collapse at \SI{16}{kyr}.
Then, radius drops to $\SI{1.7e-3}{au}$ or about \SI{3.6}{\rj} (Jovian radii).

This is an example of a clump that survives the early inward migration due to the fast accretion.
If it had stayed at its initial mass, it would have migrated into the hot inner disc long before its pre-collapse time, and would have been tidally or thermally disrupted.
Due to the increasing mass, the clump's Hill radius remains larger than \SI{3}{au} before \SI{16}{kyr}.
The disc's temperature is always lower than the clump's effective temperature, and the clump is not thermally disrupted.
However, the time until the second collapse occurs is slightly delayed through the irradiation from the disc (Sect.~\ref{sec:irrad}).
This outcome is aided by the fact that the fragment is born after the end of the infall phase, where the outer disc is colder.
Also, the fast accretion is possibly facilitated by the absence of other fragments.

The collapsed companion continues to migrate inwards on a timescale of \SI{e2}{kyr} as seen in the middle left panel.
It crosses the disc's inner edge at \SI{235}{kyr} at \SI{0.05}{au} and is considered in the model to be accreted by the primary. It could, however, in principle also become a Hot Jupiter, but we currently do not model the further evolution once a planet enters the stars magnetospheric cavity. 
The corresponding increase of the primary mass is visible in the top left panel. Observationally speaking, this system contains after the end of (main) infall phase for about 0.2 Myr a massive hot (2000 K) companion migrating through the inner 10 au of the disc.
\subsection{A system forming two fragments}\label{sec:2frag}
\begin{figure*}
    \begin{subfigure}[pt]{0.49\textwidth}
        \includegraphics[width=\linewidth]{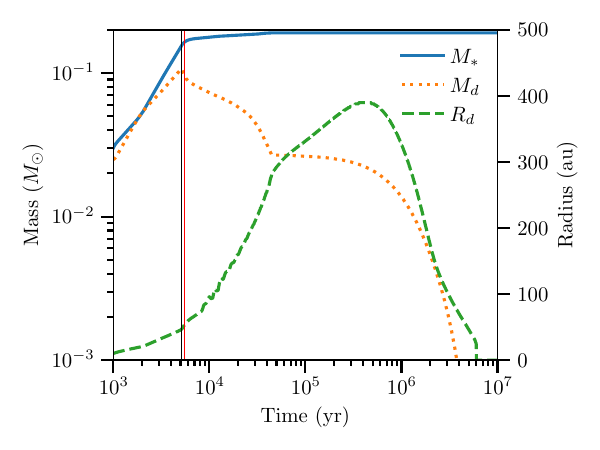}
    \end{subfigure}
    \begin{subfigure}[pt]{0.49\textwidth}
        \includegraphics[width=\linewidth]{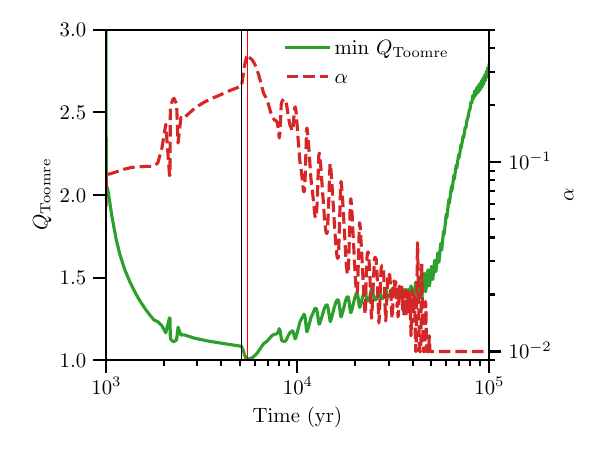}
    \end{subfigure}
    \begin{subfigure}[pt]{0.49\textwidth}
        \includegraphics[width=\linewidth]{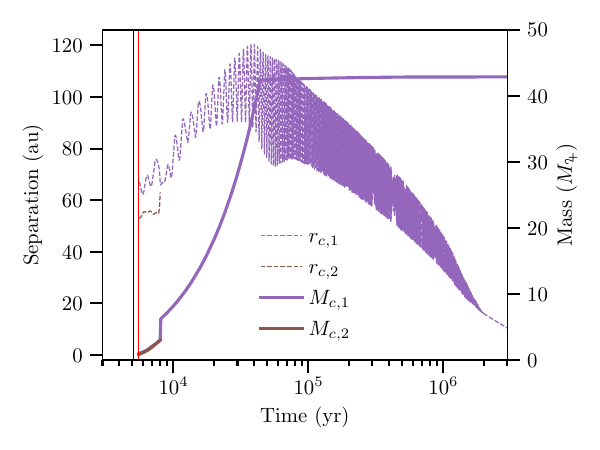}
    \end{subfigure}
    \begin{subfigure}[pt]{0.49\textwidth}
        \includegraphics[width=\linewidth]{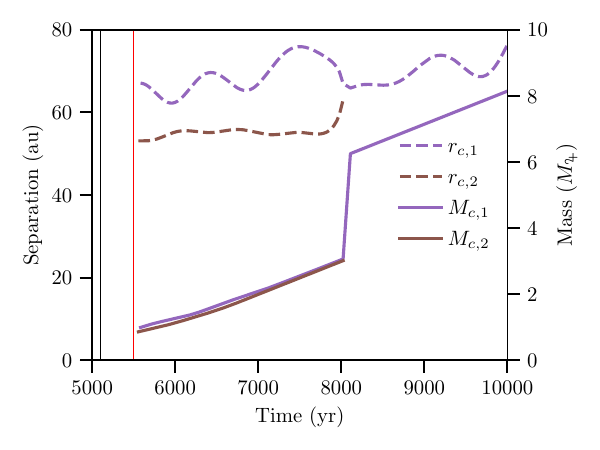}
    \end{subfigure}
    \begin{subfigure}[pt]{0.49\textwidth}
        \includegraphics[width=\linewidth]{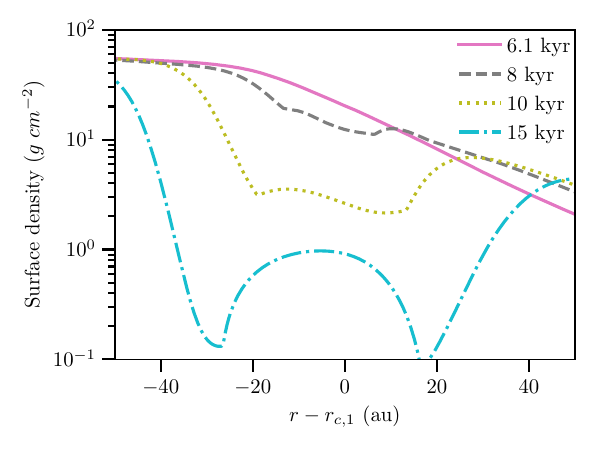}
    \end{subfigure}
    \begin{subfigure}[pt]{0.49\textwidth}
        \includegraphics[width=\linewidth]{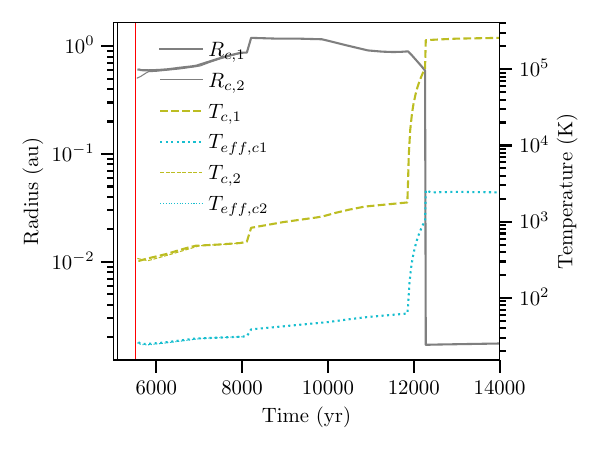}
    \end{subfigure}
    \caption{Evolution of a system yielding two fragments in a single fragmentation event.
    The panels are as in Fig.~\ref{fig:2frag}), except here, separation, mass, radius, and temperature for two companions (with subscript 1 and 2) are shown in the middle and the bottom right panels.
    The surface density shown in the bottom left panel is centered on the first companion. In the end, a massive brown dwarf companion at 5 au remains.}
    \label{fig:2frag}
\end{figure*}

Increasing the complexity further, we next discuss a system where two fragments form. Here, N-body interactions and collisions start to play a role. This system is similar to the one discussed above and is initialised with a \SI{.03}{\msun} star and a \SI{0.026}{\msun} disc. 
The disc starts less massive than the star, but accretes quickly and its mass reaches that of the star at \SI{2}{kyr}.
The star accretes faster after that, and reaches \SI{0.2}{\msun} by the end of the simulation.
The infall continues up to \SI{5.1}{kyr}, and at this time the stellar mass is \SI{0.17}{\msun} and the disc's mass is  \SI{0.11}{\msun}.
The evolution of disc and stellar mass is shown in the top left panel of Fig.~\ref{fig:2frag}.
The disc fragments at \SI{5.5}{kyr}, i.e. again shortly after the end of infall.
In contrast to the system discussed above, \textit{two} clumps form.
The evolution of $Q_\mathrm{Toomre}$ and $\alpha$ (top right panel) is qualitatively similar to that discussed before.
However, there is more oscillation in both parameters due to the more complex dynamics of the two fragments.

The middle panels of Fig.~\ref{fig:2frag} show the evolution of mass and semi-major axis of both fragments.
These clumps do not migrate inwards.
Instead, the combination of their mutual gravitational interaction and the interaction with the disc leads to a slow outward motion until the two clumps collide at \SI{8}{kyr}, i.e. shortly after their emergence.
The merged fragment with a combined mass of $\approx$~\SI{6}{\mj} migrates outwards in the increasing disc with an increasing eccentricity.
The magnitude of the torque decreases until it becomes negative and the forming companion begins migrating inwards again starting at $\approx$~\SI{50}{kyr}.
The eccentricity is damped slowly.
The bottom left panel of Fig.~\ref{fig:2frag} shows the surface density, centred on the first fragment for different times.
The gap opens less quickly compared to the system from Fig.~\ref{fig:sfrag} due to the orbit's eccentricity.
The bottom right panel depicts the evolution of temperature and radius of both fragments.
Until the collision, $R_\mathrm{c}$, $T_\mathrm{c}$, and $T_\mathrm{eff}$ are increasing slowly due to the increase in the clumps' mass. 
The curves corresponding to the second clump are in thin solid line and are hard to see since they are partly covered by the first clump's lines.
The evolution is almost the same for both clumps due to their very similar masses and ages.
After the collision, the surviving clump contracts slowly, increasing its temperature, until it undergoes second collapse at \SI{12.3}{kyr}.
While the dynamical evolution in this system is complicated due to the many factors involved, the overall behaviour of the surviving companion is characterised by an early outward migration followed by a slow inward migration.
Outward migration occurs because this companion is located at the outer edge of the disc after the interaction with the other clump.
There, the positive torque from the inner disc dominates.
The first companion continues to accrete gas, and enters the brown dwarf (BD) regime.
It carves a deep gap despite being on an eccentric orbit.
This prevents material from the outer disc from crossing its orbit and the inner disc depletes.
Once the inner disc is depleted, it no longer exerts a torque on the companion and the migration direction is inward.
The inward migration becoming slower with decreasing disc mass and stops after \SI{6}{Myr} as the disc dissolves.
The companion survives as \SI{43}{\mj} BD with a semi-major axis of \SI{5}{au}.

\subsection{A complex system with multiple fragmentation events}\label{sec:comp}
\begin{figure*}
    \begin{subfigure}[pt]{0.49\textwidth}
        \includegraphics[width=\linewidth]{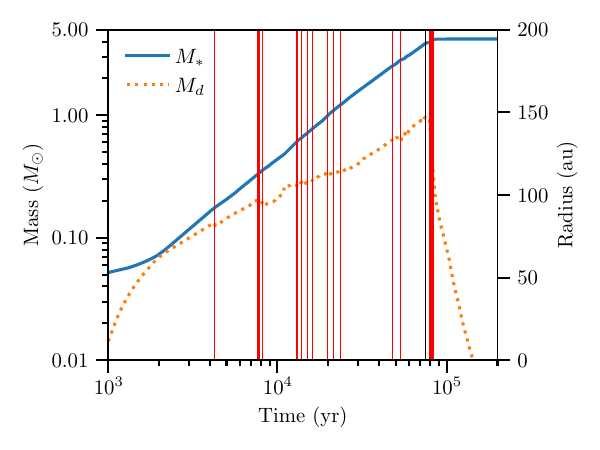}
    \end{subfigure}
    \begin{subfigure}[pt]{0.49\textwidth}
        \includegraphics[width=\linewidth]{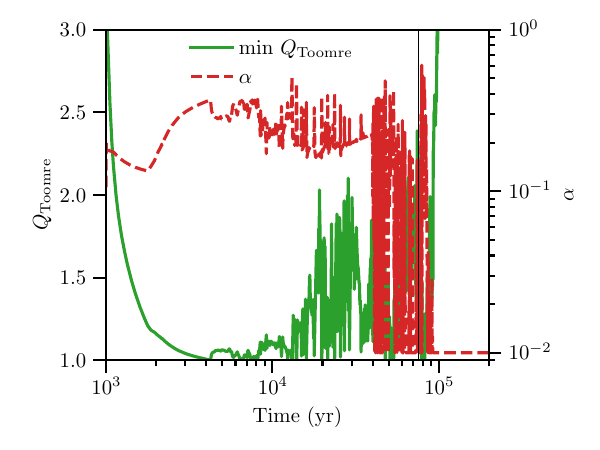}
    \end{subfigure}
    \caption{Evolution of a system with many fragments (same as Fig.~\ref{fig:nonfrag}).}
    \label{fig:mfrag}
\end{figure*}

After a fragmentation event, mass is removed from the disc, which drives it towards stability.
If the infall phase is sufficiently long, the disc can become unstable again.
Furthermore, the tidal interaction of companions with the disc can also drive it towards instability.
In this way, multiple fragmentation events, each yielding one or more fragments, are possible.
This leads to situations where many fragments are in the disc at the same time.
An example of such a system is shown in Fig.~\ref{fig:mfrag}.
The left panel shows the time evolution of the stellar and disc masses.
This is a very massive system, with the primary reaching \SI{4.2}{\msun}, the disc mass is \SI{1}{\msun} at its highest.
The infall phase lasts for \SI{75}{kyr}, which corresponds to a long duration. 
The disc fragments dozens of times and forms in total (over its lifetime) $\sim$~\num{100} fragments. 
The fragmentation events (sometimes closely spaced in time) are shown as vertical red lines in the figure.
The right panel shows the minimum of $Q_\mathrm{Toomre}$ as well as $\alpha$ as a function of time.
Both quantities show a lot of variations due to the many fragments and their interaction with the disc.
This phase lasts until \SI{100}{kyr}.
Just after the end of the infall phase, a last intense burst of fragmentation occurs.
Many clumps are scattered to the outer disc, where the majority of the mass resides.
There, they accrete much of the gas disc in a short time, as seen in the sudden drop of $M_\mathrm{d}$ in Fig.~\ref{fig:mfrag} around \SI{100}{kyr}.
This also drives the disc towards stability, with $Q_\mathrm{Toomre}$ increasing sharply and $\alpha$ dropping to its background value.

The behaviour of the companions is very complex.
Clumps interact gravitationally with each other or collide, or they might become ejected or thermally disrupted.
There are up to \num{30} objects in the disc at the same time. 

\subsection{Diversity of outcomes: system architectures, masses, semi-major axes and eccentricities}\label{sec:outcomes}
\begin{figure*}
  \includegraphics[width=\linewidth]{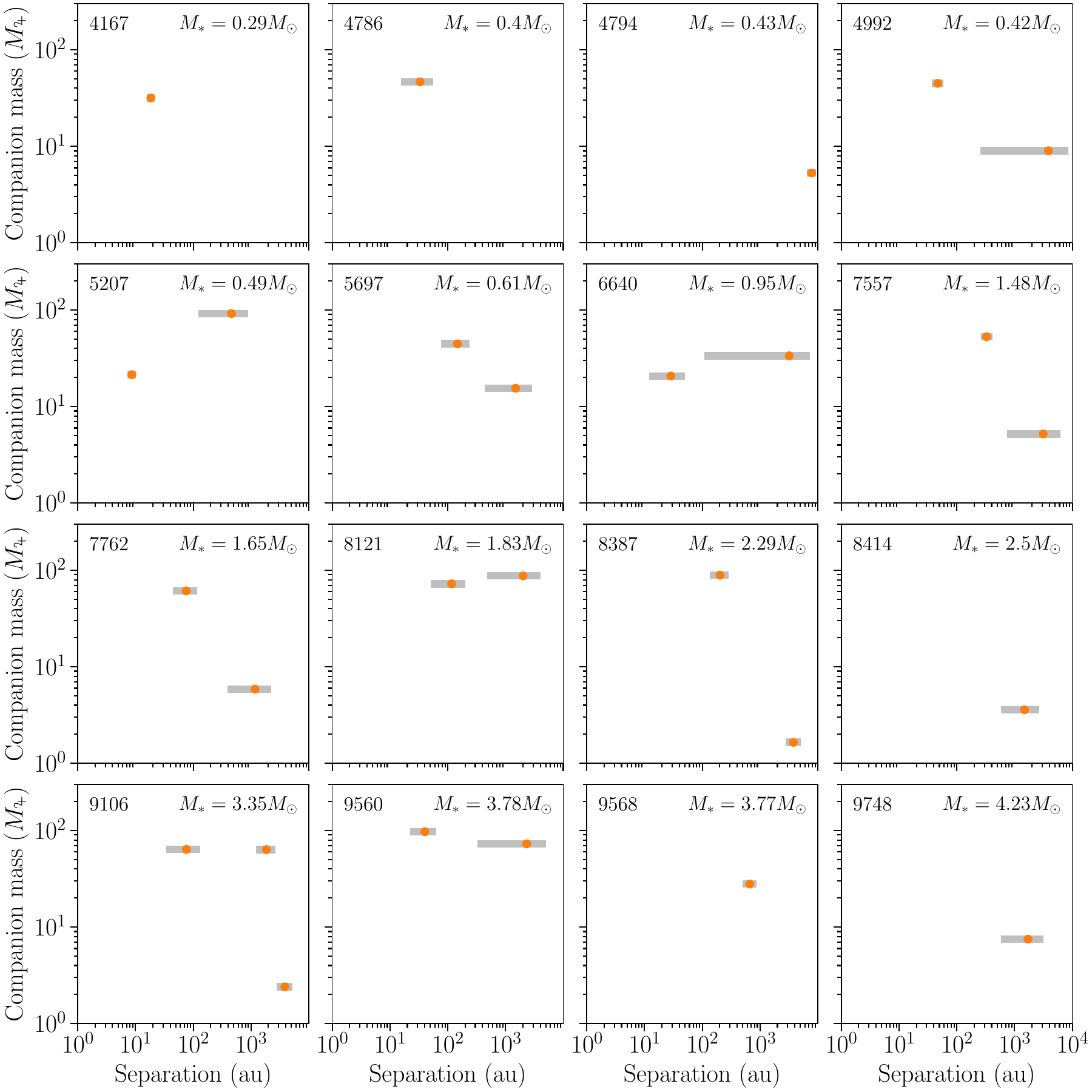}
  \caption{Mass versus semi-major axis at the end of the simulation for a selection of systems.
  The eccentricity is given as a grey bar.
  The index of the system as well as the final primary mass is given at the top of each panel.
  See text Sect.~\ref{sec:outcomes} for further explanations.}
  \label{fig:outcome}
\end{figure*}

The systems' architecture after \SI{100}{Myr} can be quite different.
We discuss these in detail on a population level in \citetalias{Schib2025b}. 
Here, we look at a number of individual systems to illustrate possible outcomes of the combined processes described above.
Surviving companions have masses from $\approx$~\SIrange{1}{200}{\mj}, semi-major axes from $\approx$~\SIrange{1}{10000}{au} and appear in single, double or triple systems.
Figure~\ref{fig:outcome} shows a collection of example systems from the baseline population DIPSY\nobreakdash-0 discussed in \citetalias{Schib2025b}.
Each panel depicts mass versus semi-major axis of the companions.
Their eccentricity is shown as grey bar.
Orbital parameters were calculated in Jacobi coordinates.
Companion masses are typically on the order of a few tens of Jovian masses as seen in  most panels of the figure.
Eccentricities can vary from close to zero up to large values.
Small eccentricities are often seen when all but one companion are lost (e.g. through collisions or gravitational interactions) early.
Any eccentricity of the remaining companion will then be damped through the interaction with the disc.
The  top left panel shows such a system.
Single companions can also have significant eccentricities though, as seen in the rightmost panels of the third and forth row.
This typically happens if a companion is scattered out of the disc or if additional companions are only lost after the disc phase.
In both cases, the companion's eccentricity would not be damped.
For systems with two surviving companions, masses can increase or decrease with semi-major axis (as seen e.g. in the second row).
The masses of the two companions can be similar (like in the lower two panels in the second column or very different (as in system 8387).
In systems with two companions, a typical configuration consists of a massive inner companion on a nearly circular orbit with a lower mass outer companion with a high eccentricity.
Examples for this are the top two systems in the rightmost row.

\section{Model limitations}\label{sect:Discussion}

\subsection{Alpha-viscosity}\label{sect:disc_alpha}
The simulations presented in this paper were run with thevalue of $\alpha = \num{e-2}$ for the  background alpha viscosity.
We did this to reproduce observed disc lifetimes (\citetalias{2021A&A...645A..43S,2023A&A...669A..31S}) and for consistency with \citetalias{Schib2025b}. This also allowed us to avoid the additional complexity that arises from a more detailed treatment of the disc's viscosity.

Here we consdider that the magneto-rotational instability (MRI, \citealt{1991ApJ...376..214B}) might not be the main driver of angular momentum in circumstellar discs \citep[e.g.][]{Lesur2023}.
A single value for $\alpha_\mathrm{BG}$ throughout the disc is probably an oversimplification independent of the value.
However, our value for $\alpha_\mathrm{BG}$ could be reasonable in the outer disc regions we are mostly interested in, if the disc is sufficiently ionised there \citep[e.g.][]{Cilibrasi2023}.
Global models of viscous discs typically apply alpha-values between \num{e-3} and \num{e-2} \citep{2016MNRAS.461.2257K,2021A&A...656A..70E}.
Using a lower value of \num{e-3} or \num{e-4} for $\alpha_\mathrm{BG}$ would lead to disc lifetimes that are  longer than the observed ones, as discussed in \citetalias{2021A&A...645A..43S} (see also \citealt{HaischJr.2001,Mamajek2009,Emsenhuber2023a}).
It may also not reproduce observed accretion rates (see Weder et al. in rev.).
The inclusion of magneto-hydrodynamic disc winds (see e.g. \citealt{2016A&A...596A..74S,2023A&A...674A.165W}) or photoevaporation models with much larger mass loss rates \citep{2019MNRAS.487..691P,2021MNRAS.508.1675E,2018MNRAS.481..452H} may lead to higher overall mass loss rates and would enable the use of a  lower $\alpha_\mathrm{BG}$ with the same disc lifetimes (i.e. with a similar long-term evolution). 
We note that the effect of wind-driven accretion on orbital migration, an important ingredient of our model, is not yet well understood \citep{2020A&A...633A...4K,2022A&A...658A..32L,WafflardFernandez2023}.
In this work, we are mostly interested in the early phases of disc formation and evolution where self-gravity plays a role. We thus kept the model simple in order to avoid additional complexity and also for consistency with \citetalias{2021A&A...645A..43S} and \citetalias{2023A&A...669A..31S}.
The choice of the background $\alpha_\mathrm{BG}$ has, however, no influence on the early evolution of our discs and their fragmentation, since the transport of angular momentum is dominated by global instabilities and spiral arms at this stage (Sect.~\ref{sec:global} and \ref{sec:spiral}), which  in our model correspond to a (significant) increase of the (total) alpha over the background value.
In Appendix~\ref{app:lv}, we tested the effect of using a lower value of \num{e-3} for $\alpha_\mathrm{BG}$.
The results shown there indicate that companions may end up on wider orbits in this case due to the slower type~II migration.
In Sect.~6.6 of \citetalias{Schib2025b}, we performed a similar test on a population level.

\subsection{Torque densities}\label{sect:disc_torque}
The torque densities we applied (\ref{sec:hmt}) were derived from 3D hydrodynamic simulations of planet-disc interaction, as discussed.
Our approach is still new and therefore there are  uncertainties regarding the migration rate and gap formation.
Here, we applied the torque densities also for companions with masses beyond the range in which they were tested.
While the results from \citet{2010ApJ...724..730D} indicate that the mass dependence is weak, this regime should be investigated in future studies. 
Maybe more importantly, we applied the torque densities also when the disc is self-gravitating.
To our knowledge, no torque densities for self-gravitating discs have been published yet, and this limitation needs to be overcome in dedicated studies.
Another important challenge is that, despite the treatment of the torque densities near the companion, gaps seem to become too deep (tend to zero) over long timescales.
While this is probably not crucial for this study (since the torque contribution from the gap bottom is very small anyway), it is of key importance for studies that include solids.
We are currently working on an improved prescription that should handle these situations.
The simulations on which the high mass torque is based are also locally isothermal.
A more comprehensive treatment of thermodynamics might not be very important for massive companions that open deep gaps.
However, some recent results \citep[e.g.][]{Ziampras2025} also indicate that thermodynamics plays a key role in the gap-opening process.
The influence of e.g. radiation transport in the parameter range of our study should also be investigated in future research. 

\subsection{Heliocentric disc model}\label{sec:disc_helio}
The basic type of disc model we use assumes that the primary is at the origin, while the disc is revolving around it.
For a disc-star system (only) this assumption is reasonable since the disc is assumed rotationally symmetric.
As long as any companion in the system is much less massive than the primary, the barycentre would remain close to the origin.
However, when studying massive companions as we do in some cases here, this assumption can become problematic.
The barycentre can shift away from the primary which leads to torques on the disc that would make it eccentric.
Furthermore, the companions' orbits would be inconsistent with the disc's dynamics. For example, a companion with a circular orbit around the barycentre no longer remains at the same location in the disc.
Such problems arise when a companion accretes at a very high rate for a prolonged period.
However, hydrodynamic simulations indicate that accretion rates onto clumps rarely exceed \SI{e-3}{\mj/yr} for any length of time.
\citep{2013ApJ...767...63S} performed 3D radiation hydrodynamics simulations of massive companions at wide separation as expected to result from fragmentation.
They find long-term accretion rates onto the proto gas-giants of up to \SI{3e-4}{\mj/yr}.
\citep{2019MNRAS.486.4398F} study the migration of massive, wide companions including gas accretion, comparing seven different hydrodynamic codes.
They do not find sustained accretion rates above \SI{e-3}{\mj/yr} also for companions with an initial mass of \SI{12}{\mj}.
We therfore limit the accretion rate onto the companions to \SI{e-3}{\mj/yr}.
We note, that accretion rates above this value were reported in \cite{2012ApJ...746..110Z}.
The accretion limit effectively restricts the applicability of our model to systems with a primary that is dominant in mass.
The study of, for example, binaries of comparable mass interacting with the disc is not possible.
We discuss this restriction further in \citetalias{Schib2025b}.

\subsection{Photoevaporation}\label{sec:disc_photo}
The prescriptions for internal and external photoevaporation (\ref{sec:disp}) are standard models that do not yet include  some of the latest results on disc photoevaporation (e.g. \citealt{2021MNRAS.508.1675E}.
The main purpose of including photoevaporation in our work is (together with the background alpha viscosity) to regulate the disc lifetimes to observed lengths (see also \citetalias{2021A&A...645A..43S,2023A&A...669A..31S}).
The exact nature of mass loss is less important, since we are mostly interested at what happens in the outer disc very early on, before photoevaporation becomes a relevant process.
However, in \citetalias{Schib2025b} we also investigate the effect of having a higher rate of external photoevaporation.

\subsection{Accretion of solids, grain growth and sedimentation}\label{sec:disc_solid}
A detailed treatment of solids (dust, pebbles) is currently missing in our model.
Solid grains inside a clump can influence its evolution, for example by changing the opacity \citep{helledbodenheimer2011, 2019MNRAS.488.4873H}.
A higher opacity would likely lead to a longer pre-collapse timescale, making the clumps' survival less likely and vice versa (see also \citealt{Baehr2019}).
On the other hand, grains could grow and sediment to form a core \citep{1998ApJ...503..923B,2006Icar..185...64H,2008Icar..195..863H,2008Icar..198..156H}, and grain growth and settling could also lead to a decrease of the clump's opacity, as discussed in \citet{helledbodenheimer2011}, and thus make it reach the second collapse earlier.
These are important aspects that we plan to address in future work.

\section{Summary and conclusions}\label{sect:Conclusions}
In this work, we  present and discuss a new global numerical model that can be used to study the formation of a star-and-disc system, including disc fragmentation, clump formation, and clump evolution as well as various interaction processes (e.g. clump-clump and clump-disc).
The model represents an improvement over existing similar models by combining several key physical effects not self-consistently linked previously.
We used a precursor of this model in previous studies \citepalias{2021A&A...645A..43S,2022A&A...664A.138S,2023A&A...669A..31S} to investigate the formation and fragmentation of circumstellar discs, as well as gas accretion and orbital migration of companions.
The new version of the model presented here combines the functionalities of previous work and adds a number of physical effects.
It can now be used to study the formation and evolution of systems of companions formed in disc instability ranging in mass from the planetary mass regime over brown dwarfs to (low) stellar masses.  By using a low-dimensional  approach (axisymmetric discs, spherically symmetric clumps and companions, but with 3D N-body interactions) the model is also suitable to perform population syntheses of objects formed in the context of disc instability.
The main addition to the model is the inclusion of one or several clumps after fragmentation and their evolution and interaction with the disc as well as with each other. With this extension, the current model consists of three main parts, described below

The first main part is the model of the circumstellar disc (Sect. \ref{sect:discmodel}). It is characterised by the following aspects:
\begin{itemize}
    \item The evolution of the gas surface density is modelled in the framework of the viscous disc picture in the 1D axisymmetric vertically integrated approximation (Sect. \ref{sec:disc}). Several source and sink terms are included: infall, disc photoevaporation, gas accretion and mass loss by the embedded companions.
    \item The viscosity of the disc is based on the $\alpha$-parametrisation (Sect. \ref{sec:visc}) and contains time-variable contributions that depend on the disc's global (Sect. \ref{sec:global}) and local stability ($Q_{\rm Toomre}$) parametrising the transport of angular momentum  by spiral arms (Sect. \ref{sec:spiral}) and a constant background contribution (Sect. \ref{sec:background}).
    \item As disc masses can be comparable to the stellar mass, the discs' auto-gravitation is taken into account in the calculation of the angular frequency and the vertical scale height (Sect. \ref{sec:auto}).
    \item As the primary mass increases strongly and its properties evolves, we use tabulated (proto)stellar evolution tracks (Sect. \ref{sec:star}), where the luminosity also includes the accretional luminosity. 
    \item Star and disc start from negligible initial masses and are then built up by infall from a molecular cloud core (Sect. \ref{sec:infall}). Infall rates are taken from hydrodynamic simulations, while infall radii are chosen to obtain observed Class 0 disc sizes. Infall durations are set to obtain the given final stellar masses.
    \item The disc's temperature structure, which (together with the gas surface density and the angular frequency) is key for the stability, is found from energy conservation in the vertical direction. This includes the contributions from viscous heating, stellar irradiation, shock heating by the infalling gas, background irradiation from the MCC, and radiative cooling at the disc's surface (Sect. \ref{sec:temperature}). 
    \item External FUV-driven and internal EUV-driven disc photoevaporation are included, as they are important for the long-term evolution post-infall and the eventual dissipation of the discs (Sect. \ref{sec:disp}). 
\end{itemize}

The second main part of the global model (Sect. \ref{sec:clump}) deals with the formation and evolution of gaseous clumps forming from the self-gravitational fragmentation of a disc. Here, the following aspects are included:
\begin{itemize}
    \item By checking the Toomre parameter and cooling criteria (different during the infall and post-infall phases) at each moment and orbital distance in the disc, we can determine whether and where the disc fragments (Sect. \ref{sect:frgmentation}).
    \item When a disc is found to fragment, one or several fragments are inserted in the fragmenting region. Their initial masses can be set according to several prescriptions found in the literature (Sect. \ref{sect:initialfragmass}). 
    \item After formation, clumps can accrete gas from the disc (Sect.~\ref{sec:acc}). Thus, we can apply the Bondi+Hill-accretion model presented in \citetalias{2022A&A...664A.138S}. This can lead to significant mass growth. The gas accreted by the clumps is removed from the disc to ensure mass conservation. 
    \item The clumps' interior evolution (cooling, contraction) is modelled by  interpolating in pre-calculated evolutionary tables of isolated clumps (Sect.~\ref{sec:tracks}). We approximately correct for the delaying effect caused by disc irradiation (Sect. \ref{sec:irrad}).
    The evolution tracks include in particular also the second collapse where the clumps turn from very extended (au-scale) low-density objects into much more compact ones (Jovian radius-scale). 
    \item Clumps can lose mass through tides if the approach close to the primary (Sect.~\ref{sec:massloss}); namely, if their physical size given by the interior evolution tracks becomes larger than the Hill sphere (Roche lobe overflow). Thermal destruction is included in this step as well.
\end{itemize}
The third main part of the global model comprises the interaction of clumps and companions with the disc and with each other.
\begin{itemize}
    \item The gravitational interaction with the gas disc is modelled by using torque densities \citepalias{2022A&A...664A.138S} to take into account the two-way exchange of angular momentum between companions and the disc (Sect.~\ref{sec:migration}).
    This leads to orbital migration and gap opening, as well as the damping of eccentricities and inclinations of the clumps.
    We have developed a new approach based on dynamical friction \citep{2020MNRAS.494.5666I} that is applicable in the context of gap-opening companions.
    \item The gravitational interaction between companions is calculated by applying an N-body integrator (Sect.~\ref{sec:nbody}). The N-body interactions (full 3D, in contrast to the other model parts ) lead to scatterings, ejections, and collisions with the host star and among the clumps and companions. 
    Collisions are treated in a simplified way and integration continues up to \SI{100}{Myr} to include potential late dynamical events. 
\end{itemize}

While it is possible to apply this new global model for disc instability population synthesis (see \citetalias{Schib2025b}), aspects of the model, such as the treatment of the infall phase or the companion-disc interaction, can also be used in general studies of companion formation via disc instability.
Following the description of the model, we showcase it (Sect. \ref{sec:sys}) by studying a number of systems with increasing complexity, studying the evolution and impact of parameters such as the primary mass and disc mass as well as mass and semi-major axis of companions.

These simulations demonstrate the complexity of modelling planet and companion formation via DI when the effects included in the model interact. 
We illustrate this by discussing in detail four simulations of increasing complexity.
The first example is a system with a disc that becomes self-gravitating, but does not fragment (Sect. \ref{sec:nonfrag}).
Then we look at a system which does fragment, but only a single fragment forms that migrates to the inner edge of the disc (\ref{sec:sfrag}).
The third example is a system in which two fragments form and later collide, with the merged companion surviving and ending up as a $\sim$40 $\mj$ brown dwarf (Sect. \ref{sec:2frag}.
Fourth, we look at a complex system that fragments multiple times with strong interaction between the dozens of fragments that emerge in the disc (Sect. \ref{sec:comp}).

Finally, as an outlook with respect to the second companion paper, we  discuss the diversity of outcomes in terms of system architectures in the case where there is at least one surviving companion by discussing the masses, semi-major axes, and eccentricities of sixteen different systems (Sect. \ref{sec:outcomes}).
Despite the complexity of the model and the many coupled physical mechanisms included, our model has some important limitations:
A fixed and high value for the background alpha-viscosity (Sect.~\ref{sect:disc_alpha}), uncertainties related to the application of torque densities (Sect.~\ref{sect:disc_torque}), the heliocentric disc model (Sect.~\ref{sec:disc_helio}), the prescriptions for photoevaporation (Sect.~\ref{sec:disc_photo}) and the absence of potential solid accretion, grain growth, and settling which are still missing in the current model (Sect.~\ref{sec:disc_solid}).

We conclude that a variety of interconnected physical processes, such as gas accretion, orbital migration, and N-body gravitational interactions, exert a strong influence on the inferred population of forming  objects. Furthermore, our findings highlight that the assumptions underlying theoretical models play a pivotal role in determining the final architecture and properties of the systems that form in such conditions. 
In \citetalias{Schib2025b} of the series, we used the new model presented here to establish a large-scale population synthesis in the disc instability model that allows us to confront the predictions of this formation paradigm with observations in a quantitative way.

\begin{acknowledgements}
We thank Lucio Mayer, Allona Vazan and Simon Grimm for the insightful discussions.
O.S. and C.M. acknowledge the support from the Swiss National Science Foundation under grant 200021\_204847 `PlanetsInTime'. R.H. acknowledges support from the Swiss National Science Foundation under grant 200020\_215634. Part of this work has been carried out within the framework of the NCCR PlanetS supported by the Swiss National Science Foundation under grants 51NF40\_182901 and 51NF40\_205606. Calculations were performed on the Horus cluster of the Division of Space Research and Planetary Sciences at the  University of Bern.
\end{acknowledgements}

\bibliographystyle{aa}
\bibliography{main}


\begin{appendix}
\section{Lower background viscosity}\label{app:lv}
\begin{figure}
    \includegraphics[width=\linewidth]{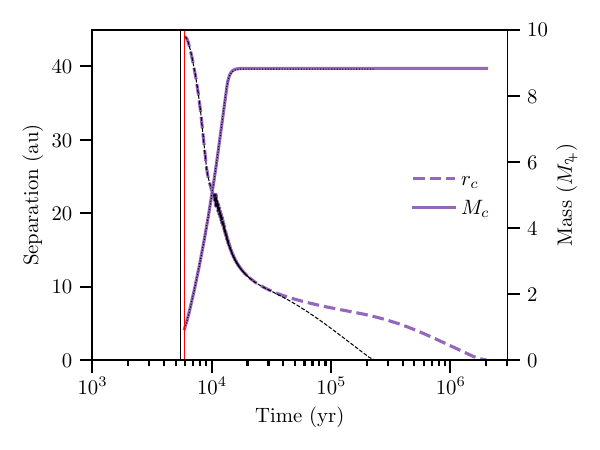}
    \caption{Mass-separation evolution of the system with one fragment at a lower background viscosity.
    The results from Sect~\ref{sec:sys} are shown in black.}
    \label{fig:msapp1}
\end{figure}

\begin{figure}
    \includegraphics[width=\linewidth]{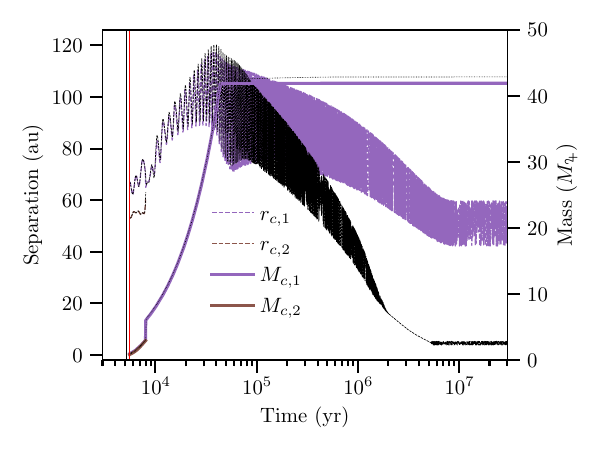}
    \caption{Mass-separation evolution of the system with two fragments at a lower background viscosity.
    The results from Sect~\ref{sec:sys} are shown in black.}
    \label{fig:msapp2}
\end{figure}

For the simulations presented in Sect.~\ref{sec:sys} we used a value of \num{e-2} for the background viscosity $\alpha_\mathrm{BG}$.
Here, we test the sensitivity of our results to this choice, by applying a lower value of \num{e-3}.
Fig.~\ref{fig:msapp1} and \ref{fig:msapp2} depict the time evolution of mass and separation for the systems discussed in Sect.~\ref{sec:sfrag} (\ref{fig:msapp1}) and \ref{sec:2frag} (\ref{fig:msapp2}).
Fig.~\ref{fig:msapp1} shows the system with a single fragment. As expected, the evolution is identical while the viscosity is determined by spiral arms (until $\approx \SI{30}{kyr}$, compare to top left panel of Fig.~\ref{fig:sfrag}).
The final masses are the same, since the mass evolution is essentially finished at this point.
The subsequent evolution of the separation is much slower at lower viscosity due to the longer type~II migration timescale. The outcome is the same, though: the companion migrates all the way to the inner edge of the disc and is accreted on the primary.
The evolution of the system with a high viscosity is shown with thin black lines for comparison (the mass is barely visible as it is unchanged).

In Fig.~\ref{fig:msapp2}, the analogous evolution for the system with two fragments is shown.
Again, the early evolution is the same and the long-term migration is much slower.
In this case, the surviving companion ends up at $\approx \SI{50}{au}$.

\section{Hill radius for high mass ratios}\label{app:rhill}
The expression for $R_\mathrm{H}$ given in Sect.~\ref{sec:acc} was derived assuming $q \equiv M_\mathrm{c}/\mstar << 1$.
It is inaccurate for $q \gtrsim 0.1$ (in fact, $R_\mathrm{H}$ is greater than \num{0.5} for $M_\mathrm{c} = \mstar$).
This is because the expression for $R_\mathrm{H}$ results from using the binomial approximation on a quintic equation in the derivation.
In order to avoid this difficulty, we use the following polynomial approximation for $q > 0.01$:
\begin{equation}
\begin{aligned}
    R_\mathrm{H} / r_\mathrm{c} & \approx 0.1171 + 3.41867 q - 31.2062 q^2 + 203.082 q^3 \\
    & - 879.93 q^4 + 
 2617.62 q^5 - 5534.93 q^6 + 8550.38 q^7 \\
    & - 9841.34 q^8 + 8546.69 q^9 - 5634.9 q^10 \\
    & + 2819.48 q^{11} - 1062.45 q^{12} + 296.371 q^{13} \\
    & - 59.3077 q^{14} + 8.04718 q^{15} - 0.662781 q^{16} \\
    & + 0.0250073 q^{17},
\end{aligned}
\end{equation}
which is accurate to $q \approx 3$.

\end{appendix}

\end{document}